\documentclass[]{interact}

\usepackage{epstopdf}
\usepackage{graphicx}
\usepackage{amsmath}
\usepackage{amssymb}
\usepackage{amsbsy}
\usepackage[nottoc]{tocbibind}

\usepackage[numbers,sort&compress]{natbib}
\bibpunct[, ]{[}{]}{,}{n}{,}{,}
\renewcommand\bibfont{\fontsize{10}{12}\selectfont}

\theoremstyle{plain}

\theoremstyle{definition}

\theoremstyle{remark}

\newcommand{\ma}[1]{\mathrm{#1}}

\newcommand{\dma}[1]{_{\mathrm{#1}}}

\begin{document}

\articletype{}

\title{Exozodiacal clouds: Hot and warm dust around main sequence stars}

\author{
\name{Quentin Kral\textsuperscript{a}\thanks{CONTACT: Q. Kral, Email: qkral@ast.cam.ac.uk},Alexander V. Krivov\textsuperscript{b}, Denis Defr\`ere\textsuperscript{c}, Rik van Lieshout\textsuperscript{a}, Amy Bonsor\textsuperscript{a}, 
Jean-Charles Augereau\textsuperscript{d,e}, Philippe Th\'ebault\textsuperscript{f}, Steve Ertel\textsuperscript{g}, J\'er\'emy Lebreton\textsuperscript{d,e}, and Olivier Absil\textsuperscript{c}}
\affil{\textsuperscript{a} Institute of Astronomy, University of Cambridge, Madingley Road, Cambridge CB3 0HA, UK; \textsuperscript{b} Astrophysikalisches Institut und Universit\"{a}tssternwarte, Friedrich-Schiller-Universit\"{a}t Jena, Schillerg\"{a}{\ss}chen 2-3, D-07745 Jena, Germany;
\textsuperscript{c} Space sciences, Technologies, and Astrophysics Research (STAR) Institute, Universit\'e de Li\`ege, 19c all\'ee du Six Ao\^ ut, B-4000 Li\`ege, Belgium; 
\textsuperscript{d} Institut de Planetologie et d’Astrophysique de Grenoble (IPAG, UMR 5274), Univ. Grenoble Alpes, F-38000 Grenoble, France;  
\textsuperscript{e} CNRS, Institut de Planetologie et d’Astrophysique de Grenoble (IPAG, UMR 5274), F-38000 Grenoble, France;
\textsuperscript{f} LESIA-Observatoire de Paris, UPMC Univ. Paris 06, Univ. Paris-Diderot, 92195 Meudon, France;
\textsuperscript{g} Steward Observatory, Department of Astronomy, University of Arizona, 933 N. Cherry Ave, Tucson, AZ 85721, USA}
}


\maketitle

\begin{abstract}
A warm/hot dust component (at temperature $>$ 300K) has been detected around $\sim$20\% of A, F, G, K stars.
This component is called ``exozodiacal dust'' as it presents similarities with the
zodiacal dust detected in our Solar System, even though its physical properties and
spatial distribution can be significantly different. Understanding the origin and evolution
of this dust is of crucial importance, not only because its presence could hamper future
detections of Earth-like planets in their habitable zones, but also because it can provide
invaluable information about the inner regions of planetary systems.
In this review, we present a detailed overview of the observational techniques used in the detection 
and characterisation of exozodiacal dust clouds (``exozodis'') and
the results they have yielded so far, in particular regarding the incidence rate of exozodis
as a function of crucial parameters such as stellar type and age, or the presence of an outer cold debris disc.
We also present the important constraints that have been obtained, on dust size distribution
and spatial location, by using state-of-the-art radiation transfer models on some of these systems.
Finally, we investigate the crucial issue of how to explain the presence of
exozodiacal dust around so many stars (regardless of their ages) despite
the fact that such dust so close to its host star should disappear rapidly due to the coupled
effect of collisions and stellar radiation forces. Several potential mechanisms have been
proposed to solve this paradox and are reviewed in detail in this paper. The review finishes by presenting
the future of this growing field.
\end{abstract}


\section{The basics of exozodiacal dust}

The term exozodiacal dust (short exozodi) is used here to refer to warm or hot dust (with $T>300$K) orbiting around a main sequence star. The zodiacal dust in our Solar System is part of this category. However, exozodis can be much brighter
and located at different radial locations than the zodiacal dust. Exozodis are to be distinguished from their colder counterparts, called debris discs, for which the observed dust is produced
by quasi steady state collisions in belts (similar to the Kuiper belt) composed of planetesimals and large rocky bodies orbiting at tens of au \citep{2008ARA&A..46..339W,2010RAA....10..383K}.

\begin{figure}
\centering
\includegraphics[scale=0.4]{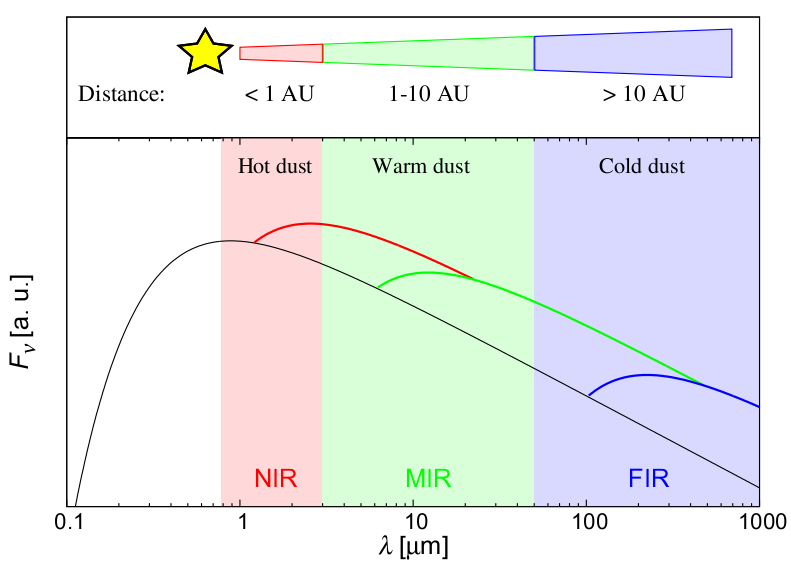}
\caption{Illustration of the hot (red), warm (green) and cold (blue) belts in terms of their respective contributions to different parts of the SED \citep{Kirch17}. The stellar photosphere is shown as a black line. The height of the excesses are here in arbitrary units [a. u.].}
\label{fig:kirsch2}
\end{figure}

Two populations of exozodis are observed in the infrared (IR): warm and hot exozodis. In the remainder of the paper, we will make a distinction between warm exozodis that we detect in the mid-IR (specifically at about 10$\mu$m, where habitable zone dust peaks) and hot exozodis that we detect in the near-IR (in the H or K-band). 
Note that these two populations can co-exist and this is not a physical boundary but rather an observational one (see Fig.~\ref{fig:kirsch2}).
Space-based observatories such as Spitzer or WISE searched for warm dust but its presence is found to be rare at the limited sensitivity (due to calibration accuracy and uncertainties on the predicted photosphere) 
of these observations \citep[$\sim$ 1\% for young $<$120 Myr systems,][]{2013MNRAS.433.2334K}.  Only the brightest warm exozodis can be detected with photometry against the stellar photosphere and hot exozodis are harder to detect because the photosphere
is brighter in the near-IR. To detect exozodis, one needs to separate the stellar emission from the dust emission.  
The high spatial resolution required can only be reached with interferometry. The first detection of hot dust was achieved around Vega \citep{2006A&A...452..237A}, and then
around $\tau$ ceti \citep{2007A&A...475..243D} using the CHARA array interferometer at K-band. It is found that this hot dust (or observationally speaking, an excess of order 1\% relative to the photosphere) is rather common \citep[$\gtrsim$ 10\%,][]{2014AA...570A.128E}.

The presence of such high levels of dust in the inner regions of planetary systems, particularly in old ($>$100Myr) systems is rather puzzling. The lifetime of dust against collisions so close to its host star is short. Quantitatively, a steady state belt between 0.3 and 0.5au 
whose dust mass is $10^{-6}$M$_\oplus$ (a typical mass to produce the observed excesses), which is composed of big rocky bodies colliding and creating hot dust through a collisional cascade, km-sized bodies in that belt
would survive only $\sim$ 100yr owing to large collisional rates \citep{2007ApJ...658..569W}. Therefore, in the standard models, it is hard to explain the observed dust in steady-state as being created in situ. In order to explain the rather ubiquitous presence of exozodis, a scenario that produces 
dust at a rate of $\sim 10^{-9}$M$_\oplus$/yr is required \citep{defrere2011} or a mechanism that increases the dust lifetime by orders of magnitude \citep[e.g.][]{2014A&A...571A..51V}.
Such scenarios have been proposed and will be detailed in section \ref{origin}. At the moment, none can provide a full explanation for the whole observed sample. The full explanation may lie in a combination of mechanisms.

Exozodis highlight the presence of large quantities of warm dust in the habitable zone (HZ) of planetary systems and could prevent us from detecting the holy grail, an exo-Earth: a planet similar to Earth in mass and size orbiting inside the habitable zone of a sun-like star. 
It is critical to characterise the frequency and level of dust in the habitable zones of nearby stars in time for surveys that will directly image Earth-like planets. Indeed,
\citet{2012PASP..124..799R} show that direct imaging techniques trying to image exo-Earths could be affected by tiny amounts of dust at levels of our own zodiacal dust, which is already 1000 times fainter than most exozodis that we observe today. 
For coronagraphic missions, \citet{2015ApJ...808..149S} show how the number of planets that can be studied decreases with the exozodiacal dust level around nearby main-sequence stars. For interferometric observations, \citet{2010A&A...509A...9D}
show that the tolerable exozodi density not to hamper exo-Earth detections is $\sim$ 15 times the density of the zodiacal cloud. Also, small dust clouds whose IR-excesses could not be detected by current instruments can mimic an Earth-like astrometric signal which may affect
future astrometric missions looking for exo-Earths \citep{2016A&A...592A..39K}.

Exozodis may be considered as a source of noise for exoplanet detections but they are also a powerful way to probe the inner regions of planetary systems. Indeed, the very simple fact that hot dust is present in these systems constrains the planetary system
architecture. If planets are located in this warm/hot dust region, resonant structures such as the Earth resonant ring should be present \citep{1999ASPC..177..374W}. Their detection would be an indirect way to probe the presence of close-in planets
as done for colder belts further away \citep[e.g. $\beta$ Pic,][]{2001A&A...370..447A}.

The aim of this review is to summarise our current observational and theoretical knowledge and understanding of exozodiacal dust. We use the term exozodiacal dust to refer to the presence of hot and warm dust ($>$ 300K) in inner regions of planetary systems. The system with the 
best understood zodiacal dust is our own Solar System, and we firstly review the current state-of-the-art observations in Sect.~\ref{zodi}. This is contrasted with our current knowledge of exozodiacal dust discs, summarised in Sect.~\ref{obscon}. Of critical importance is the impact of the presence of 
exozodiacal dust on the potential detection of exo-Earth planets, discussed in Sect.~\ref{hamper}. The observational properties of exozodis can be used to infer the dust properties. Sect.~\ref{grater} summarises the state-of-the-art radiative transfer modelling and what this tells us 
regarding the properties of exozodis. An overview of the plausible origins of these exozodis is given in Sect.~\ref{origin}. The review finishes with a discussion of the potential growth of the field in the future, and how currently planned instruments will provide many of the insights 
required to complete our understanding of exozodiacal dust.

\section{The Solar System Zodi}\label{zodi}

Before discussing exozodiacal dust around other stars, it is natural
to have a look at the zodiacal dust in our own Solar System
(Fig.~\ref{fig:zodi_sketch}).
Here, we give a brief overview of what is known about dust in the region interior
to Jupiter's orbit (i.e., at distances $r \le 5$au). 
For more details on the Solar System's interplanetary dust cloud, the reader is referred
to dedicated reviews \citep[e.g.][]{dohnanyi-1978,leinert-gruen-1990,gruen-et-al-2001,mann-et-al-2006}.

\subsection{Observation methods}\label{inzoobs}
Information about dust in the inner Solar System comes
from a variety of remote sensing and
in situ observations. These include measurements of zodiacal light brightness and polarisation, thermal 
emission observations in the infrared, 
microcrater counts on lunar samples, direct in situ detections by dust detectors aboard spacecraft,
collecting particles in the stratosphere and in near-Earth orbit,
analyses of deep sea and polar ice sediments, statistics of radar and visual meteors, and others. 
Each of the methods listed above is sensitive to dust grains of different sizes and constrains different 
grain properties.
For instance, remote sensing mostly probes micrometer-sized and larger particles,
whereas in situ detections effectively trace fine, submicrometer-sized dust.
Depending on the method, one can deduce or constrain the dust density, the size or mass
distribution of grains, their velocities and  orbits, optical properties, morphology,
mineralogy, and even electric charges they acquire in the solar radiation and plasma environment.
The majority of the methods works best to constrain dust at 1au.
Information on dust at other distances derives from observations of the zodi's scattered light and thermal emission 
and from in situ spacecraft detections. Knowledge of the dust environment close to the Sun, at $r < 0.3$au,
is the poorest and comes mostly from the brightness observations
of the solar Fraunhofer corona (a.k.a. F-corona; sunlight scattered by near-solar dust in the forward direction) made 
during solar total eclipses or from satellites equipped with coronagraphs.

\subsection{Basic facts}
The infrared luminosity of the zodiacal dust cloud amounts to $\sim 10^{-7}$ of the solar 
bolometric luminosity \citep{nesvorny-et-al-2010,2012PASP..124..799R}, and is of the 
same order as that of the Kuiper belt dust in the outer Solar System \citep{vitense-et-al-2012}.
The dust emission peaks at $\lambda \sim 19\,\mu$m, which translates to a mean temperature of 
$\sim 270$~K \citep{nesvorny-et-al-2010}.
The total mass of the dust cloud inside Jupiter's orbit is estimated to be
between $\sim 10^{-9}$ and $\sim 10^{-8}$ M$_\oplus$ \citep{fixsen-dwek-2002,nesvorny-et-al-2010}.
The mass density of interplanetary dust close to the Earth's orbit
is roughly comparable to that of the solar wind \citep{mann-et-al-2010}.

\begin{figure*}
\centering
\includegraphics[width=0.80\textwidth,angle=0]{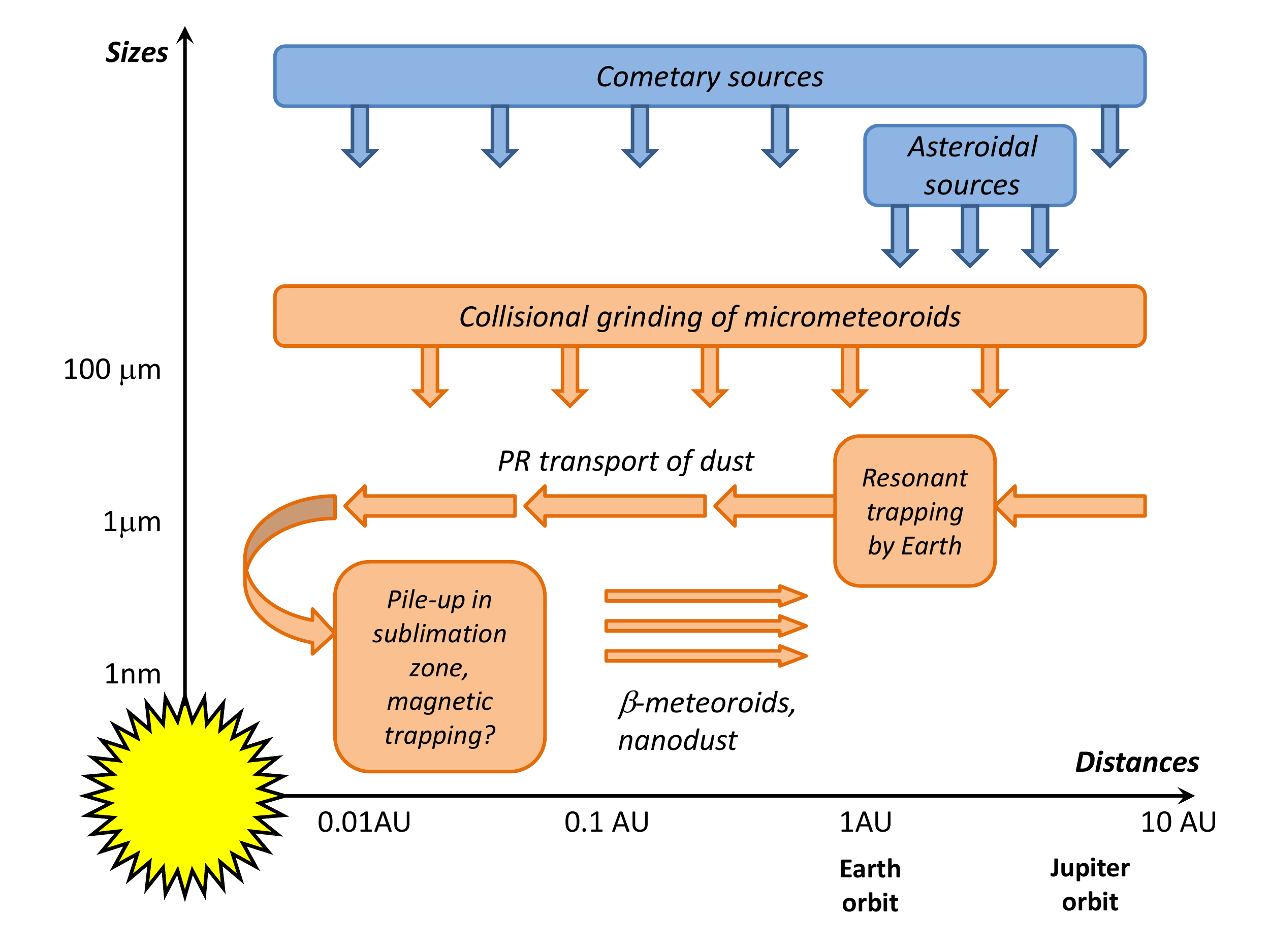}
\caption{
Schematic of the Solar System's zodiacal cloud in the size--distance 
plane (top: parent bodies, bottom: fine dust, right: Jupiter's orbit, left: dust sublimation 
zone). 
The sketch illustrates the
main physical processes operating in the cloud,
the zones of enhanced concentration of 
micrometer- and submicrometer-sized dust,
and radial transport of dust. 
\label{fig:zodi_sketch}
}
\end{figure*}

\subsection{Radial distribution}\label{radialzd}
Dust pervades the interplanetary space at all heliocentric
distances, down to a few solar radii (see Fig.~\ref{fig:zodi_sketch}).
The number density roughly scales with distance as $r^{-1.3}$ \citep[e.g.][]{gruen-et-al-1985}.
This is slightly steeper than the reciprocal of distance, which would be expected
for dust that is produced by sources in nearly-circular orbits outside the inner cloud and
transported inward by the Poynting-Robertson effect and the solar wind drag \citep{briggs-1962}.
Including collisional losses would only make the slope flatter, as would the assumption of parent bodies
moving in elliptic orbits \citep{gorkavyi-et-al-1997b}. The only way to get a slope steeper than a 
reciprocal of distance is to assume that the source of visible dust is extended
\citep{leinert-et-al-1983}.
This source is probably a population of directly invisible larger meteoroids, which are spread
over the cloud and produce smaller dust grains in mutual collisions.
On top of the average density profile discussed above,
there is an enhancement of the dust density around the Earth's orbit.
This has been found based on IRAS observations and interpreted as trapping of dust in mean-motion 
resonances with the Earth \citep{dermott-et-al-1994}. Recently, a subtle dust density enhancement has also been detected around the orbit of Venus \citep{2013Sci...342..960J}.
Closer in, \citet{peterson-1967} and \citet{macqueen-1968} inferred a dust concentration at $\sim 4$ solar 
radii, i.e. just outside the distance where dust is expected to sublimate. The properties and a possible
origin of this circumsolar ring have been subject to 
intensive observational and theoretical studies 
\citep[e.g.][]{mukai-yamamoto-1979,lamy-et-al-1992,mann-1992,kimura-et-al-1997b,%
krivov-et-al-1998b,mann-et-al-1999,2008Icar..195..871K}, but remain a matter of debate.


\subsection{Size distribution}
\begin{figure}
\centering
\includegraphics[width=0.55\textwidth,angle=0]{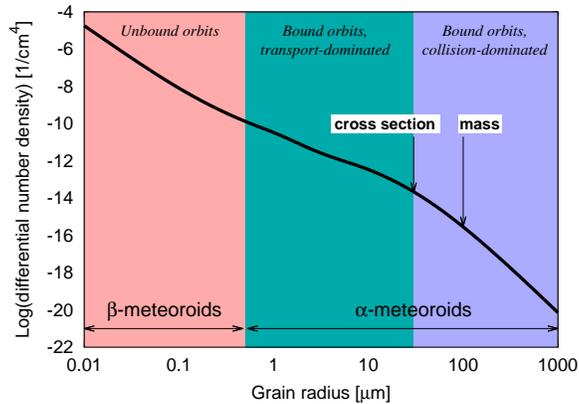}
\caption{
Size distribution of dust at $1$au. Curve is the \citet{gruen-et-al-1985} empirical model, based on 
the measurements. Coloured areas correspond to dynamical regimes expected from the theory: grains on unbound 
orbits (salmon), grains on bound orbits drifting inward without experiencing frequent collisions (green), and grains on 
bound orbits with  collisional lifetimes shorter than the transport timescales (purple). Sizes at which the slope of 
the distribution changes seems  to roughly reflect transitions between different regimes. Labelled are sizes of grains 
that dominate the cross section and mass distributions of the dust complex.
\label{fig:gruen}
}
\end{figure}

The overall size distribution (Fig.~\ref{fig:gruen}) is best known at 1au \citep[see, e.g.][]{gruen-et-al-1985}.
Inside and outside 1au the data are scarcer,
although various models exist and are in use by NASA and ESA
\citep[e.g.][]{divine-1993,staubach-et-al-1997,kelsall-et-al-1998,dikarev-et-al-2004,mcnamara-et-al-2004}.
The differential size distribution slope of particles larger than $\sim 100\,\mu$m was found to be steeper
than $-4$, so that the mass density is dominated by particles $\sim 100\,\mu$m in radius.
Most of the contribution to the cross section comes from somewhat smaller grains, $\sim 30\,\mu$m in size
\citep{gruen-et-al-1985,love-brownlee-1993}.
This is approximately the size at which the Poynting-Robertson drift timescale becomes comparable to the
collisional lifetimes of dust grains. Particles smaller than $\sim 30\,\mu$m typically drift toward 
the Sun without being further disrupted by collisions.
Their size distribution slope is flatter than $-3$ down to about a micrometer.
Below that size, which corresponds roughly to the radiation pressure blowout limit, the distribution
gets steeper than $-4$ again.

\subsection{Submicrometer- and nanometer-sized dust}\label{submzd}
There is ample evidence for the presence of copious amount of grains of submicrometer sizes, being expelled 
outward from the Sun in hyperbolic orbits by radiation pressure. These particles, often called
$\beta$-meteoroids \citep{zook-berg-1975}, have been detected by dust detectors aboard a number of 
spacecrafts both in and outside the ecliptic plane \citep[e.g.,][and references therein]{wehry-mann-1998}. 
They come predominantly from the solar direction and 
probably originate in close proximity to the Sun, perhaps in the innermost dust concentration zone 
described above.
They may be either the tiniest debris from meteoroid collisions or sublimation remnants of larger particles 
brought inward by drag forces. 
The tiniest, nanometer-sized particles streaming from the vicinity of the Sun at roughly the solar 
wind speeds have also been detected 
with the plasma instrument aboard the STEREO spacecraft \citep{meyervernet-et-al-2009}.
Some of these nanoparticles may be trapped by the solar magnetic field inside 
about 0.2au \citep{2010ApJ...714...89C}.

\subsection{Mineralogy and morphology}

These properties have been probed by laboratory analyses of interplanetary dust particles (IDPs)
collected in the stratosphere or by Earth satellites \citep{brownlee-1979}.
IDPs are found to be complex mixtures of thousands or millions of mineral grains
and amorphous components with a rich, mostly chondritic, 
elemental composition \citep{jessberger-et-al-2001}. 
Optically, they appear as opaque, dark objects, with absorbing materials including
carbon, sulfides, and GEMS (glass with embedded metals and sulfides).
Their properties vary broadly from one IDP to another.
Some of the IDPs are porous, fluffy aggregates with low densities ($\sim 1$~g~cm$^{-3}$)
and anhydrous mineralogy;
these have been interpreted as being of cometary origin.
Others are nearly compact and smooth, have higher densities ($\sim 3$~g~cm$^{-3}$) and include
hydrated minerals; these must have been produced by asteroids.
There is also a strong correlation between these properties and
atmopheric entry speeds, which are typically $>18$~km~s$^{-1}$ for ``cometary'' and
$<14$~km~s$^{-1}$ for ``asteroidal'' IDPs,  consistent with higher orbital eccentricities and
inclinations of comets as compared to asteroids \citep[e.g.,][]{joswiak-et-al-2007}.
Some support for that ``dichotomy'' comes from a direct comparison of IDPs with return samples,
such as those of the Startdust mission to the comet P81/Wild2  \citep{brownlee-et-al-2006},
the Hayabusa mission to the asteroid Itokawa \citep{Tsuchiyama,nakamura-et-al-2012}
and, most recently, from the suite of experiments onboard the Rosetta mission to the
comet 67/P Churyumov-Gerasimenko \citep[e.g.][]{Fulle}.


\subsection{Sources}
While it is obvious that both comets and asteroids make non-negligible contributions
to the dust supply, there has been controversy as to which of these two classes
of small bodies is the dominant source of dust.
Based on dynamical models, it has been argued that the asteroidal dust may prevail in the cloud 
\citep{dermott-et-al-1984,sykes-et-al-1988, dermott-et-al-1992}.
However, there is now a growing bulk of evidence that most of the dust in the inner Solar System
comes from short-period comets. 
Notably, the vertical extent of the zodi is broader than expected from asteroids,
but it matches nicely the latitudinal distribution of Jupiter-family comets
\citep{nesvorny-et-al-2010}.
Also, a recent analysis of the leading/trailing brightness asymmetry of the dust ring around
the Earth's orbit, as observed with the AKARI satellite, suggests that millimeter-sized cometary dust
grains mainly account for the zodiacal light in the infrared, while
the contribution of asteroidal dust is less than $\sim 10$ per cent \citep{Ueda}.
The cloud may either be fed by disintegration of cometary nuclei into tails of large fragments
with their subsequent collisional grinding
\citep{sykes-et-al-1986,reach-et-al-2000,nesvorny-et-al-2011} or
by a direct injection of dust from cometary sublimation \citep{Keller}.
If the comets are indeed the dominant dust source, one would have to explain why
atmospheric collections contain comparable amounts of asteroidal and cometary type IDPs.
This may be caused by observational biases. 
For example, cometary dust particles encounter the Earth at higher relative velocities than 
asteroidal ones. As a result, they undergo a weaker gravitational focusing which decreases 
their accretion rate to Earth \citep{flynn-1994b}.
Besides, higher atmospheric entry velocities lead to 
a stronger thermal alteration of cometary IDPs, making it more difficult to identify
them as being of cometary origin \citep{jessberger-et-al-2001}.

The view of the Solar System's inner zodi presented above offers a natural reference for a discussion
of exozodis in the subsequent sections. While zodis around other stars do not have to be similar
to ours, the Solar System's case may provide useful guidelines for interpretation of observations.
Important lessons from the Solar System are, for instance, that 1) comets may be important as dust 
sources, producing most of the dust through disruptions/splitting rather than sublimation;
2) that zones of enhanced dust density may exist close to the star;
and 3) that submicrometer dust grains may be abundantly present, with their dynamics being largely 
driven by interactions with the stellar magnetic field.

\section{Observations of Exozodis}\label{obscon}

We first provide Table~\ref{tab1} that lists all infrared intereferometers that have been used to observe exozodis.
We then introduce the resolved interferometric observations and results for the hot dust detected in the near-IR (subsection \ref{nir}) and continue with the warm dust detected in the mid-IR (subsection \ref{sec:mir}). We then provide a list of results that were obtained
from unresolved observations and the analysis of spectral features (subsection \ref{unres}). Finally, we explain and quantify how these exozodis might affect future missions looking for exo-Earths (subsection \ref{hamper}).


\begin{table}[!t]
\caption{List of infrared interferometers used for exozodiacal disc observations. From left to right, the table gives the instrument name, its waveband, its date of operation (from first fringes to decommissioning), its observing mode (V: visibility, N: nulling), the number of published  stars, and the main corresponding reference(s).}
\begin{tabular}{l c c c c l}
\hline
\hline
Instrument name & Band & Date & Mode & Stars & References \\
\hline
IOTA/IONIC & H  & 1993-2006 & V & 1 & \citet{defrere2011} \\

VLTI/PIONIER & H  & 2010-\phantom{9999} & V & 144 & \citet{2014AA...570A.128E,Marion} \\%

PTI &  K & 1995-2008 & V  & 1 & \citet{2001ApJ...559.1147C}\\

VLTI/VINCI  &  K & 2001-2003 & V  & 1 & \citet{2009ApJ...704..150A} \\

CHARA/FLUOR &  K & 2002-\phantom{9999} & V & 75 & \citet{2013AA...555A.104A,nunez} \\%

VLTI/GRAVITY &  K & 2016-\phantom{9999} & V & 1 & Defr\`ere et al. in prep \\

PFN &  K & 2009-2015 & N & 1 & \citet{2011ApJ...743..178M} \\

VLTI/MIDI & N & 2002-2015  & V & 4 & \citet{Smith:2009,2012MNRAS.422.2560S}  \\

MMT/BLINC & N & 2004-2009  & N & 6 & \citet{2009ApJ...693.1500L,Stock:2010} \\

KIN & N & 2007-2011 & N & 47 & \citet{Mennesson:2014} \\

LBTI/NOMIC & N & 2012-\phantom{9999} & N & 26 & \citet{Defrere:2015}, Ertel et al. in prep \\
\hline
\label{tab1}
\end{tabular}
\end{table}

\subsection{Resolved observations in the near-infrared}\label{nir}

Spectro-photometry cannot detect the signature of hot dust emitting below the 1\% stellar level around nearby stars because of a lack of photometric accuracy, as well as the poor accuracy of stellar photosphere models. The only way is to resolve the dust from the star. This requires
high resolution and high contrast that are reached using high-precision instruments on interferometers such as CHARA/FLUOR \citep{2003SPIE.4838..280C} or VLTI/PIONIER \citep{2011A&A...535A..67L}. 

\begin{figure}
   \centering
   \includegraphics[width=8.cm]{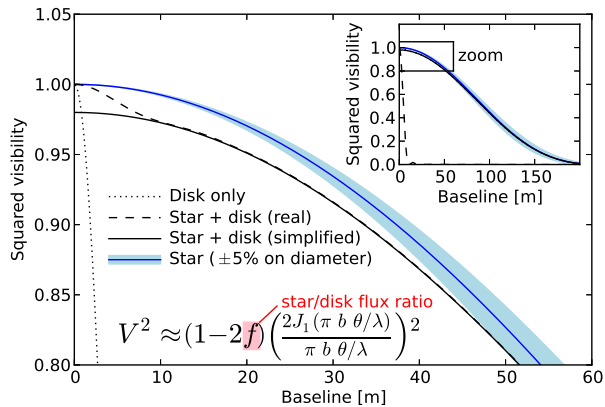}
   \caption{\label{figdetect} Illustration of the detection method of exozodis in the near-IR. Squared visibility is shown as a function of baseline for a disc only (dotted), a real (dashed) or a simplified (solid) \{disc+star\}. The real profile assumes
a uniform disc for both the star and the exozodiacal dust and a disc-to-star flux ratio of $f$=0.01 (or $f_{\rm CS}$ in the main text) whilst the simplified profile makes use of the same assumptions but uses Eq.~\ref{visib2} for the stellar model. For more details see \citet{2014AA...570A.128E}.}
\end{figure}

The observed dust in the near-IR is emitting at the 1\% stellar level in H/K-band within $\sim$ 1'' field-of-view (a radius of $\sim$ 400mas for FLUOR and 200mas for PIONIER). The observed dust can only be resolved with interferometry as it is within a few au (which corresponds to $\sim$0.1'' for a 10pc distant star). 
Using baselines of a few tens of meters in the near-IR, the extended emission of the dust
can be fully resolved, whilst the star itself remains largely unresolved. The presence of this dust around the star creates a small deficit in the measured squared visibilities compared to what is expected for a star alone owing to the addition of incoherent flux from dust. 
This deficit is then used to infer the presence of a dust cloud and enables the measurement of the disc-to-star flux ratio (see Fig.~\ref{figdetect} for more details). 

The measured visibility $V$ representing the \{star+disc\} system, assuming that the disc is uniform and fills the whole field-of-view, can be expressed as a function of baseline length $b$ \citep[see][]{2007A&A...475..243D}

\begin{equation}
\label{visib}
V^2(b)=(1-2f_{\rm CS}) \, V^2_\star(b),
\end{equation}

\noindent where $V_\star$ is the visibility of the stellar photosphere alone and $f_{\rm CS} \lesssim 1$\% is the the flux ratio between the integrated circumstellar emission within the field-of-view and the stellar photospheric emission. For a simplified star model assuming a uniform circular star, $V^2_\star$
can be expressed as follow

\begin{equation}
\label{visib2}
V^2_\star(b)=\left( \frac{2 J_1(\pi b \theta_\star / \lambda)}{\pi b \theta_\star / \lambda}\right)^2,
\end{equation}

\noindent where $J_1$ is the Bessel function of first order, $\theta_\star$ is the star's angular diameter and $\lambda$ the wavelength. In reality, more complex stellar models are used to take account of limb darkening \citep[e.g., see Eq.~6 in][]{2008A&A...487.1041A}.

One should note that $f_{\rm CS}$ is on the order of $\sim 1$\%, thus a relatively high accuracy is needed to measure $V^2$ but also to estimate the squared visibility of the photosphere $V^2_\star$. The former condition can be met with high-precision near-IR interferometers (such as FLUOR and PIONIER) and the latter is ensured
by measuring the system at short baselines, where the star is mostly unresolved so that the stellar visibility depends weakly on the photospheric model.

In terms of sensitivity, the state-of-the-art PIONIER instrument on the VLTI can detect exozodis that are at least $\sim$ 500 times brighter than the zodiacal dust. The first detection of an exozodi was around Vega 
with an excess of $1.29 \pm 0.19$\% relative to the photospheric emission using CHARA/FLUOR \citep{2006A&A...452..237A}, closely followed by another detection around $\tau$ Ceti \citep{2007A&A...475..243D}. The 
first detection with PIONIER was around $\beta$ Pic \citep{2012A&A...546L...9D}.

After these successes, two surveys in the near-IR have led to major advances in the field. The first survey was undertaken by CHARA/FLUOR in the K-band \citep{2013AA...555A.104A}, followed by a second one with PIONIER in the H-band \citep{2014AA...570A.128E}. 
We note that in addition to the wavelength difference, FLUOR was mostly targeting systems in the North whilst PIONIER observed in the South, which makes these surveys complementary.
FLUOR can only use two telescopes at a time, which limited the overall observational efficiency and led to the observation of only 43 stars observed in approximately 98 allocated nights spread over 7 years. On the other hand, PIONIER can use four 1.8m telescopes simultaneously 
and could target 92 stars in 12 nights. Using more than two telescopes enables closure phases of the detected systems to be measured. This is a measure of the deviation of the brightness distribution of an observed target from point symmetry, which can lead to disentangling
an extended emission from a companion \citep[non-zero closure phases,][]{2014A&A...570A.127M} with dust emission.
 
The results of the merged \{FLUOR+PIONIER\} sample is presented in \citet{2014AA...570A.128E}. The median sensitivity reached by PIONIER is $2.5 \times 10^{-3}$ (1$\sigma$) on the disc-to-star contrast.
This large sample ($\sim$ 125 stars) is useful to make a significant statistical analysis and find out whether these excesses are tied to any of the system's parameters such as stellar type,
age of the system, the presence of a cold debris disc, etc. This can in turn be used to put constraints on the mechanism(s) creating these hot excesses.

\begin{figure}
   \centering
   \includegraphics[width=8.cm]{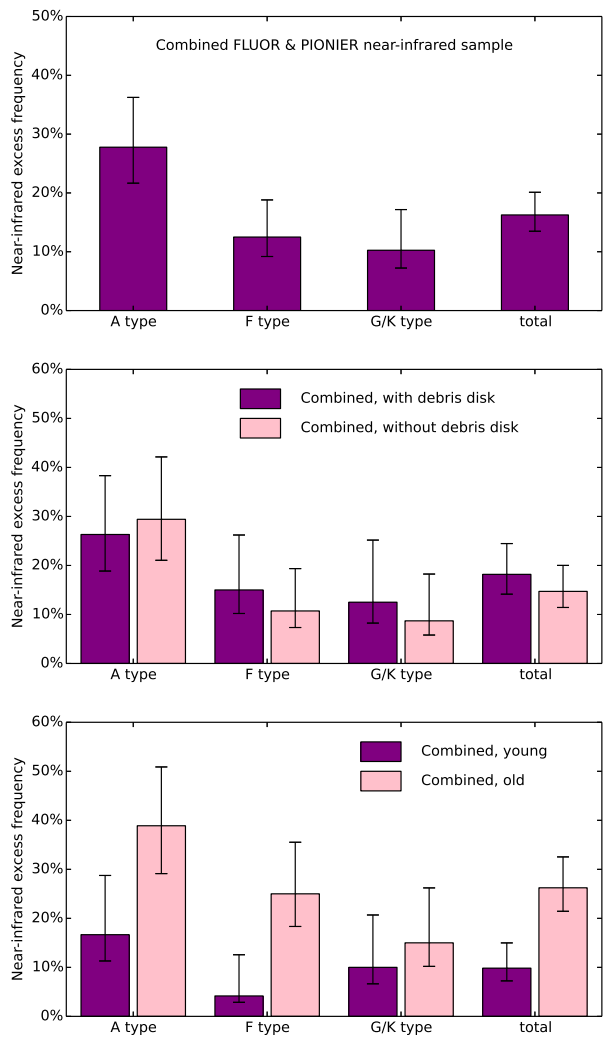}
   \caption{\label{figsteve} Statistics on the combined FLUOR+PIONIER near-IR sample \citep{2014AA...570A.128E}.}
\end{figure}

Fig.~\ref{figsteve} shows the main results from the merged sample. In this figure, the near-IR excess frequency (i.e. a significant excess above the stellar photosphere) is shown as a function of spectral type of stars (A, F, G/K). From top to bottom, it shows the combined sample (top), 
the sample separated between systems with debris discs or without (middle), and a cut as a function of age (bottom). These three plots show the main correlations that were found in this study. 
On the top plot, there is tentative evidence that the detection rate decreases with stellar spectral type from $\sim$28\% for A-type stars to $\sim$10\% for G and K type stars, which is similar to the cold debris disc trend with spectral type \citep[e.g.][]{2005ApJ...620.1010R,2013A&A...555A..11E,2014prpl.conf..521M}. 
This suggests a common origin for both phenomena. It may well be correlated to the fact that more massive solid bodies form around more massive stars due to a larger initial protoplanetary disc mass around earlier type stars \citep[$M_{\rm disc} \sim 0.01 M_\star$,][]{2011ARA&A..49...67W}. However,
this tentative trend with spectral type does not seem to be confirmed by the most recent results \citep[see][]{Marion}.

From the middle plot, we can see that there is no obvious correlation between near-IR excesses and the presence of cold dust. This is a rather surprising result as the hot dust cannot be created in situ, and is, thus, expected to be connected to an outer reservoir from which mass is transferred. 
Thus, it seems to suggest that the two processes might not have the same origin (see section \ref{origin}). However, given 1) the statistical uncertainties, 2) that we are only able to detect the brightest end of the exozodi luminosity function ($\sim$ 1000 times more luminous than our zodiacal cloud) and most luminous debris discs 
($10$-$10^4$ times brighter than our Kuiper belt), 3) that mechanisms to transfer mass inwards from an outer belt are not all explored; there is still some room for a correlation to be hidden. Indeed, it could be that there are efficient ways to
inject sufficient dust in the inner regions from a very faint Kuiper-belt-like not detectable debris disc \citep[e.g.][]{2016arXiv161102196F}. Or rather, it could mean that their presence is correlated to the architecture of planetary systems (which is generally not known). A new study by
\citet{Marion} suggests that there is, however, a correlation between the presence of warm dust (presented in more details in the next subsection) and near-IR excesses. This new correlation may favour scenarios where the hot dust is created from warm dust (e.g with PR-drag or 
planet-scattering, see Sect.~\ref{origin}).

\citet{2014AA...570A.128E} also find (bottom plot) that there is tentative evidence that detection rates increase with stellar ages. This is also surprising because if the hot dust is produced from cold belt reservoirs,
one expects these outer reservoirs to deplete collisionally over time as there is a continual mass loss through the collisional cascade \citep[e.g.][]{2006A&A...455..509K,2007ApJ...663..365W,2013A&A...558A.121K}. This is the opposite of what would be expected if it were a steady state process. This could point towards stochastic
phenomena that happen at late ages due to planets \citep[e.g.][]{2016arXiv161102196F}. However, current observations find no correlation between the presence of planets and detection rates.

For targets observed with PIONIER, the spectral slope of the H-band flux ratio can be derived using three different spectral channels. A positive spectral slope would tend to show that the flux is dominated by thermal emission, whilst a rather constant slope would point towards emission dominated
by scattered light. It is found that the spectral slope is more or less flat for most targets. Only two sources have a better fit with a black body spectrum rather than a flat slope, maybe because these exozodis have a lower temperature, $\sim$1000K instead of 2000K or because these other flat spectra 
originate from scattering and these two from thermal emission. 
Thus, it seems that there is a large diversity of exozodi architectures 
rather than the dust being right at the sublimation radius for all systems. This flat spectral slope also means that scattered light could be a significant contributor to the total 
emission \citep[e.g.][for $\beta$ Pic]{2012A&A...546L...9D} so that the observed dust can be farther than previously thought and closer to the habitable zone. To check this hypothesis, \citet{2016ApJ...825..124M} used optical polarisation measurements to estimate the amount of scattered light coming
from exozodis. They do not find any strong polarised emission around exozodis, which suggest it is consistent with thermal emission but they cannot rule out scattered light for contrived scenarios (e.g. dust in a spherical shell or face-on discs). 
It thus favours the near-IR excess to be related to dust thermal emission, meaning that the dust must be very close-in and hot.

\citet{2016A&A...595A..44E} confirmed statistically that the observed excesses persist over timescales of a few years (for most detections) and searched for variability among these targets. They find that HD 7788 is a strong candidate for variability. 
The excess around this star was first detected in 2012 and seemed to disappear for about a year before being redetected again a year later. The variability also means that a given source should be observed at different epochs to ensure detection. 
It is still too early to infer strong conclusions on the dust production mechanism.


\subsection{Resolved observations in the mid-infrared}\label{sec:mir}

Mid-infrared (8-12\,$\mu$m) observations are ideal to probe warm dust located in the habitable zone of stars at temperatures similar to that of Earth (i.e., $\sim$300\,K). As described in the introduction, characterisation of these dust discs is particularly important in the context of planetary system 
science and rocky exoplanet direct observations. While the first detections of warm exozodiacal dust around nearby main-sequence stars were achieved by spectro-photometry with space-based single-dish telescopes (see subsection \ref{unres}), these instruments do not provide the required spatial resolution to localise 
the dust grains in the region where physical and dynamical
processes need to be studied ($<$5-10\,au). In addition, they can only detect the brightest systems due to their limited sensitivity. High-contrast interferometry provides a way to go beyond both limitations by spatially resolving fainter warm excess emissions. 

\begin{figure*}
\centering
\includegraphics[height=5.1cm]{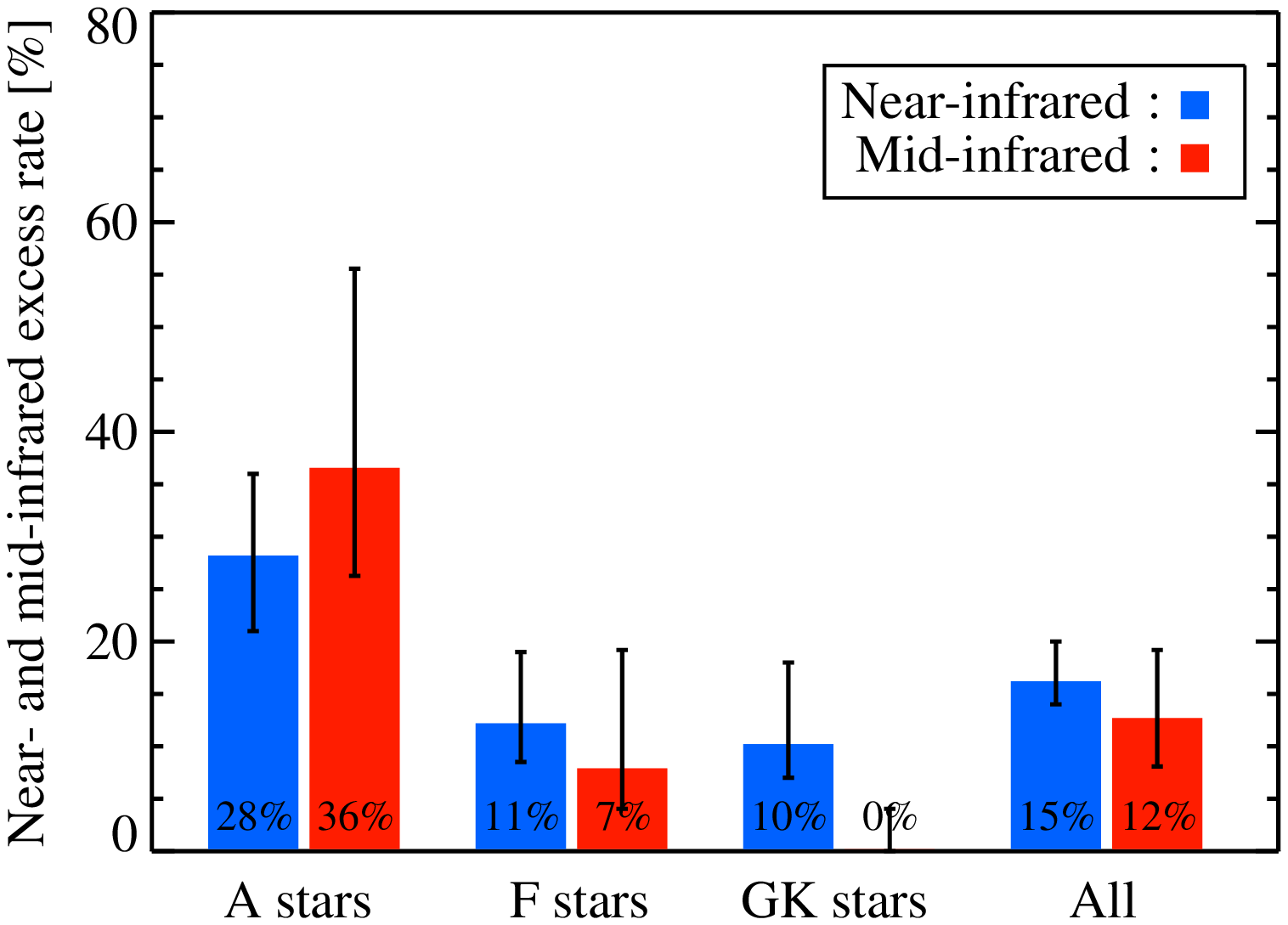}
\includegraphics[height=5.1cm]{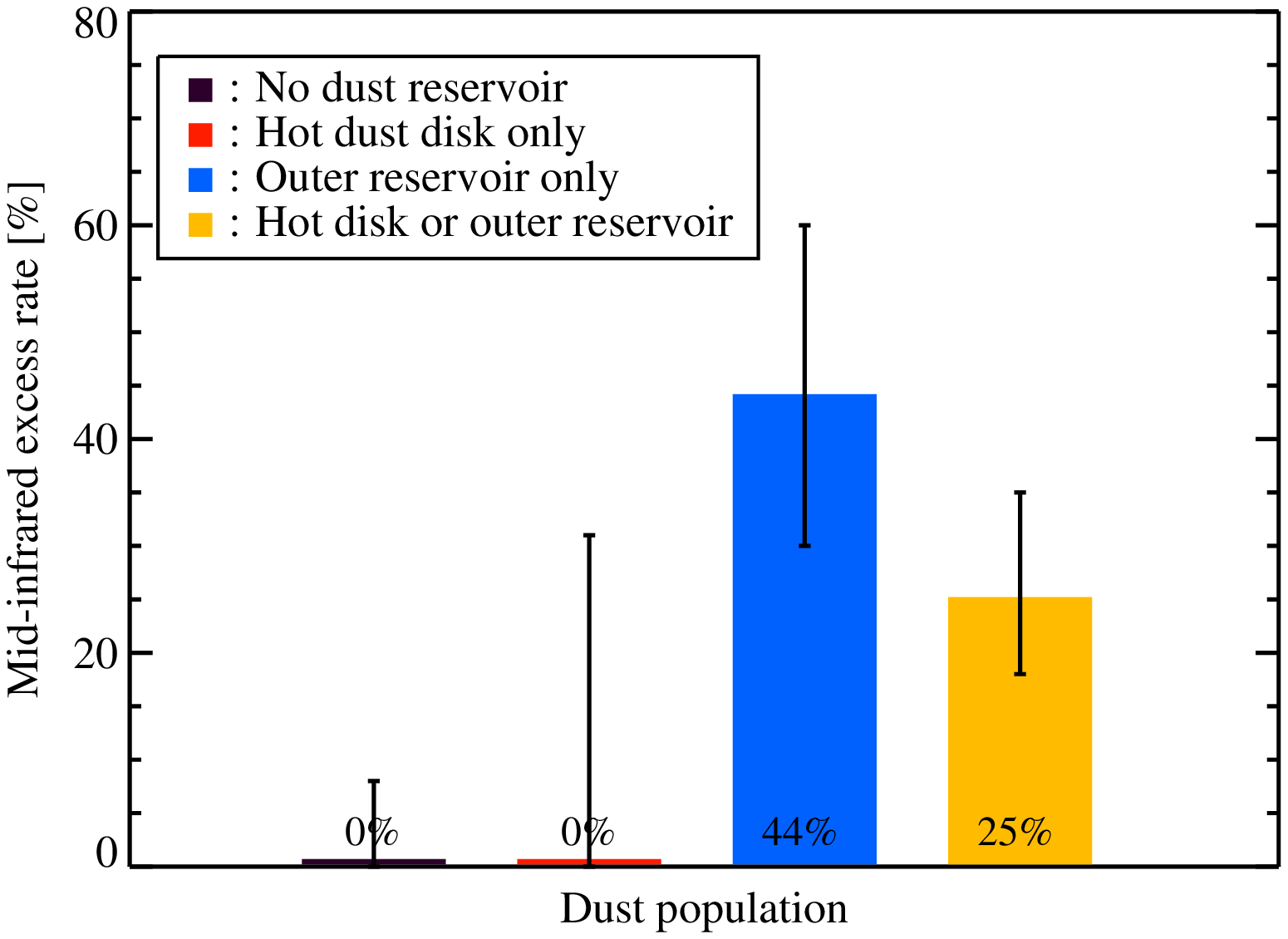}
\caption{Left: Infrared excess occurrence rate observed for different spectral types in the near-infrared (blue) and mid-infrared (red) by high-precision interferometers (data from \citet{2014AA...570A.128E} and \citet{Mennesson:2014}). For both wavelength regimes, the 
occurrence rate decreases for late-type stars, which is similar to the behaviour observed with single-dish space-based telescopes at longer wavelengths (e.g., Spitzer at 24\,$\mu$m and Herschel at 70\,$\mu$m ). Right: Mid-infrared excess rates measured by the 
KIN (8-9\,$\mu$m) for stars with various types of infrared excess previously known (or lack of it). Surprisingly, there is no correlation with stars that show hot dust detected by high-precision near-infrared interferometers. Stars having an outer dust reservoir detected at 
FIR wavelengths (70\,$\mu$m  or longer) show however a strong correlation, which suggests a physical link. }
\label{fig:MIR-results}
\end{figure*}

While some instruments such as VLTI/MIDI use a technique similar to that described in the previous section \citep[e.g.,][]{Smith:2009}, most results were obtained using nulling interferometry, a technique first proposed by \citet{Bracewell:1978} to image extra-solar planets. 
The basic principle is to combine the beams from different telescopes in phase opposition in order to strongly reduce the on-axis starlight while transmitting the flux of off-axis sources located at angular spacings given by odd multiples of 0.5$\lambda/B$ (where $B$ is the 
distance between the telescope centres and $\lambda$ is the wavelength of observation). The high-angular resolution information on the observed object is then encoded in the null depth, which is defined as the ratio of the flux measured in destructive interference and that 
measured in constructive interference. The advantage of obtaining null depth measurements is that they are more robust against systematic errors than visibility measurements and hence lead to a better accuracy \citep[e.g.,][]{Colavita:2010}.

While the first detection of warm dust by mid-infrared nulling interferometry were achieved by MMT/BLINC \citep{Stock:2010}, the first statistically meaningful survey was achieved by the Keck Interferometer Nuller \citep[KIN,][]{Millan-gabet:2011,Mennesson:2014}. A total of 47 stars 
have been surveyed by the KIN between 2008 and 2011, including 40 stars with no known companions within the instrument field-of-view (i.e. 200 mas outer working angle). At the sensivity of the KIN (i.e. 150 zodi at 1$\sigma$), only five stars showed an 8-9\,$\mu$m excess, while a marginal 11$\mu$m excess was also detected around Fomalhaut. Although the statistical 
significance of this result is limited by the sample size, it is interesting to note that excesses were only detected around main-sequence stars with types earlier than F2 (see Fig.~\ref{fig:MIR-results}, left). Interestingly, a statistical analysis of the whole data set shows a good correlation 
between the level of warm dust emission measured at 8-9\,$\mu$m and the presence of (Kuiper-belt-like) cold dust (see Figure~\ref{fig:MIR-results}, right). While warm dust detected by the KIN necessarily resides within the inner few au, this result suggests that it finds its origin in the outer 
regions of the system rather than through in situ collisions of large parent bodies. This is broadly consistent with a scenario where an outer planetesimal belt is feeding dust to the inner system through the balanced effects of PR-drag and grain collisions (see subsection \ref{prdr}), which predicts that 
the level of warm dust emission is fairly insensitive to the properties of the outer disc, as long as it is bright enough. These cold excess stars constitute prime science targets for higher precision/larger FOV mid-infrared observations of exozodiacal dust, but they are probably bad targets 
for future missions designed to directly image Earth analogues. Finally, as mentioned earlier, \citet{Marion} suggest a possible connection between warm and hot dust populations, which supports the theory that hot dust originates in material transport from an external reservoir of dust. 
However, this correlation is not present for all stars as shown by the KIN results (see Fig.~\ref{fig:MIR-results}, right). This may be due to an observational bias (i.e warm dust being too faint to be detected,) or call for large amounts of very hot grains piling up close to the sublimation radius, 
small enough to have a low mid-infrared emissivity and remain undetected by the KIN \citep[see e.g., Vega,][]{defrere2011}.

Regarding the occurrence of warm exozodis around stars with no previously detected far infrared excess (i.e, no outer dust reservoir), \citet{Mennesson:2014} derived that the median dust level has to be below 60 times the solar value with high confidence (95\%, assuming a 
log-normal luminosity distribution). To go beyond this state-of-the-art exozodi sensitivity, NASA has funded the Large Binocular Telescope Interferometer (LBTI) to carry out a survey in the N' band (9.81 - 12.41\,$\mu$m) on 40 to 50 carefully chosen nearby main-sequence 
stars \citep{2015ApJS..216...24W}. The first-light observations were obtained on a previously-known bright disc around the nearby main-sequence star $\eta$\,Crv and show an excess emission of 4.40\% $\pm$ 0.35\% over a field-of-view of 140\,mas in radius \citep{Defrere:2015}. 
This relatively low null was unexpected given the total disc-to-star flux ratio measured by \emph{Spitzer/IRS} ($\sim$23\% across the N' band), suggesting that a significant fraction of the dust lies within the central nulled response of the LBTI (79\,mas or 1.4\,au). More recently, 
the LBTI demonstrated a calibrated null accuracy of 0.05\% over a three-hour long observing sequence on the bright nearby A3V star $\beta$ Leo \citep{Defrere:2016}. This is equivalent to an exozodi density of 15 to 30~zodi for a Sun-like star located at 10\,pc, depending 
on the adopted disc model. This result sets a new record for high-contrast mid-infrared interferometric imaging and opens a new window on the study of planetary systems.

\subsection{Unresolved observations}\label{unres}
High-precision photometry with space-based single-dish telescopes provided the first detections of warm exozodiacal dust around nearby main-sequence stars. In particular, the surveys performed by the Infrared Astronomical Satellite (IRAS), 
the Infrared Space Observatory (ISO), the \emph{Spitzer Space Telescope}, and WISE have greatly improved our understanding of the incidence of warm exozodis at the bright end of the luminosity function. Initial estimates based on 
IRAS \citep{Mannings:1998,Fajardo:2000} and ISO \citep{Laureijs:2002} observations were that $\le$\,2\% of systems have detectable discs at 10\,$\mu$m. At the sensitivity of Spitzer and WISE (i.e., $\sim$1000 times brighter than our own  zodiacal cloud), 
exozodiacal dust discs around mature stars also appear to be very rare. Spitzer/IRS observations showed that only 1.0\%$\pm$0.7\% have excesses at 8.5-12\,$\mu$m \citep{Lawler:2009}. This occurence rate increases to 11.8\% $\pm$ 2.4\% of nearby, 
solar-type stars at 30-34\,$\mu$m in accordance with theoretical models \citep{2007ApJ...658..569W}. Age is an important factor as shown by \citet{2013MNRAS.433.2334K} based on WISE observations, but not the sole factor. Using a simple in situ evolution 
model to interpret the observed luminosity function, it appears that neither a picture where all systems have initially massive discs, nor one where all excesses are due to randomly timed collisions, can explain the fact that warm dust is observed around both young 
and old stars. A simple combination of these two in situ scenarios reproduces the observed luminosity function. However, it is not possible to rule out that massive warm dust discs are in fact dominated by comet delivery. 

Regarding the composition of these massive discs, mineralogical features have been observed in some warm dust systems and gave invaluable constraints on the dust properties \citep[e.g. in $\beta$ Pic,][]{2007ApJ...666..466C}. For instance, by studying the IRS 
spectrum of HD 69830, \citet{2005ApJ...626.1061B} find that prominent features of crystalline silicates such as fosterite are present. In addition, for the features to be so prominent, they claim that the grains must be of submicron size. For the same system, \citet{2007ApJ...658..584L} 
find that the dust lacks amorphous pyroxenes, PAHs and is depleted in iron and sulfure compared to solar and may originate from the disruption of P or D-type asteroids. $\eta$ Corvi (see also subsection \ref{Stype}) is another such system with strong spectral features showing 
evidence for warm water- and carbon-rich dust at $\sim$3au as well as primitive cometary material and impact produced silica \citep{2012ApJ...747...93L}. The results coming from the Spitzer spectrograph for debris discs are summarised in \citet{2014ApJS..211...25C} and \citet{2015ApJ...798...87M}.

Finally, regarding the impact on future exo-Earth imaging missions, photometric observations do not provide enough precision to draw a firm conclusion. Extrapolating over several orders of magnitudes, \citet[][]{2013MNRAS.433.2334K} suggest that at least 10\% of Gyr-old 
main-sequence stars may have sufficient exozodiacal dust to hamper the direct detection of Earth-like planets but observations at much fainter levels are required to confirm this conclusion (see Section \ref{sec:mir}). This is discussed more generally in the next subsection \ref{hamper}.

\subsection{Impact of exozodis on rocky exoplanet detection and characterisation}\label{hamper}

Another important motivation to studying hot and warm dust populations is related to the direct observation of Earth-like extrasolar planets in their habitable zones (exo-Earths) located in the habitable zone of nearby Sun-like stars. 
For instance, in the Solar System, the zodiacal disc is the most luminous component (after the Sun itself) at 
visible and infrared wavelengths \citep{kelsall-et-al-1998} and the Earth might clearly appear as an embedded clump in it for an external observer \citep{kelsall-et-al-1998}. Similarly, exozodiacal dust can act both as a source of noise and a source of confusion that must be taken 
into account in the design of future instruments that will directly characterise exo-Earths. Several independent studies have addressed this issue and concluded that visible to mid-infrared direct detection of exo-Earths would be seriously hampered in the 
presence of dust discs 10 to 20 times brighter than the solar system zodi assuming a smooth brightness distribution \citep[e.g.,][]{Beichman:2006b,2012PASP..124..799R}. 

More recently, \citet{2014ApJ...795..122S} investigated the impact of the median dust density levels on the exo-Earth candidate yield of 4-m aperture coronagraphic telescope for various exozodi luminosity distributions and different assumptions on the 
knowledge of these distributions. The results show that the exo-Earth candidate yield is divided by a factor 2 if the median exozodi level increases by a factor 10. Similarly, \citet{2012SPIE.8442E..0MD} used collisional disc models to investigate the impact of 
clumps on the detection of exo-Earths and show that the detection of planets becomes challenging beyond a density of approximately 20 zodis. Similar conclusions were obtained for mid-IR interferometers \citep{2010A&A...509A...9D}.
For astrometric missions, \citet{2016A&A...592A..39K} show that small dust clouds in inner regions of planetary systems whose IR-excesses cannot be detected by current instruments could mimic the astrometric signal of
an Earth-like planet, which may affect future astrometric missions looking for exo-Earths.
However, the prevalence of exozodiacal dust at such a level in the terrestrial planet region of nearby planetary systems is currently poorly constrained and must be determined to design these future space-based instruments. 


\section{Basic properties of exozodis}\label{grater}

We attribute the near- and mid-infrared excesses detected by
interferometry to exozodiacal dust grains located within the field of
view of the instruments (typically a few au for nearby stars). The
main characteristics of detected exozodis come from the modelling of
their spectral energy distribution, together with some modest spatial
constraints. This step is important to make the connection between the
observations (Sect.~\ref{obscon}) and the dynamical simulations
(Sect.~\ref{origin}). The modelling provides the dust properties, the
spatial distribution and the mass, which are key in order to
understand the dust dynamics and the replenishment processes. Several
systems with near- and mid-IR excesses have been modelled in
details. It appears that the resolved emissions with near-IR
interferometers can usually be attributed to submicron-sized grains
much smaller than the blow-out size. These grains are carbon-rich and
tend to accumulate near to the sublimation region, which is a striking
comparison with the hot component of the zodiacal cloud in the Solar
System, namely the F-corona (see subsection \ref{inzoobs}). The mid-IR excesses, on the other hand,
point to warm dust belts at a few au from the star and that better
compare to the zodiacal dust cloud next to Earth (see subsection \ref{radialzd}).

\subsection{Radiative transfer modelling}\label{radjc}

The simplest approach is to model the emission from an exozodi using 
a blackbody or modified blackbody to the near- and/or mid-IR excesses,
together with upper limits derived from aperture photometry
(e.g. AKARI, WISE) and/or moderate resolution spectroscopy (usually
Spitzer-IRS spectra). This yields a fractional luminosity for the
exozodi and a mean dust temperature that can be converted into a mean
distance to the star. This method provides, however, limited information regarding the nature, location and amount of exozodiacal dust. There
are also important uncertainties due to the blackbody assumption,
especially for the small grains which emit less efficiently than
blackbodies and are, therefore, hotter.

To improve our understanding of these systems, more sophisticated radiative transfer modelling
is necessary. Fortunately, the dust density is low enough to make the
exozodis optically thin, which simplifies the approach
considerably. Here, we describe more specifically the way it is
implemented into the GRaTer code which was used to model various
exozodis in details \citep[e.g. Vega, Fomalhaut and $\eta$ Corvi,
  by][respectively]{2006A&A...452..237A, 2013A&A...555A.146L,
  2016ApJ...817..165L}.


\subsubsection{Grain population and composition}

The exozodiacal dust emission is modelled assuming a population of
dust grains at a distance $r$ from the star and of size $s$ between
$s\dma{min}$ and $s\dma{max}$. Because each grain has its own
equilibrium temperature, the sublimation distance is not unique but is
a function of the grain size, making the sublimation zone a region
that can extend between 0.1 and 0.5au for an A-type star like Vega,
and between 0.01 and 0.02au (i.e. a few stellar radii) for a
solar-type star like $\tau$~Ceti. As a consequence, the differential
size distribution $\ma{d}n(r,s)$, usually assumed to be a power law
($\ma{d}n(r,s)\propto s^{\kappa} \ma{d}s$), can be truncated because
of the sublimation, and therefore depends on the distance to the star
$r$.

At a given observing wavelength $\lambda$, the total flux of an
exozodi is the sum of the light scattered off the grains plus the
thermal emission from this dust population, given by:
\begin{eqnarray}
  \Phi(\lambda,r) & = & \int_{s\dma{min}}^{s\dma{max}}
  F_\star(\lambda)\frac{f(\lambda, \varphi,
    s)\sigma\dma{sca}(\lambda,r,s)}{r^2} \ma{d} n(r,s) \nonumber \\ &
  + & \int_{s\dma{min}}^{s\dma{max}} \pi B_{\nu}\left(T\dma{d}(s,r)\right)
  \frac{\sigma\dma{abs}(\lambda,r,s)}{4\pi d_\star^2} \ma{d} n(r,s)
\end{eqnarray}
where $F_\star(\lambda)$ is the stellar flux, $d_{\star}$ the distance
to Earth and $B_{\nu}$ the Planck function. The dust optical
properties are calculated using the Mie theory valid for hard
spheres. It provides the scattering cross section ($\sigma\dma{sca}$),
the absorption/emission cross section ($\sigma\dma{abs}$), the
anisotropic scattering phase function ($f$) at scattering angle
$\varphi$, and allows to calculate the equilibrium temperature of a
grain with the star ($T\dma{d}$). For simplicity, isotropic scattering
is assumed ($f = 1/4\pi$). In practice, thermal emission dominates
over scattering for the hot exozodis in the near-IR and beyond.

The assumed composition is important in this context. In the Solar
System, both the asteroidal and cometary dust particles
incorporate significant fractions of silicates and carbonaceous
materials, and so are expected the exozodiacal dust
particles. Mixtures of silicates and carbonaceous species with various
abundance ratios are therefore considered. Because silicates sublimate
at lower temperatures than carbonaceous species, there exists a regime
where the grains are assumed silicates-free. The volume fraction of
silicates is then replaced by vacuum, leading in this model to porous
carbonaceous grains in the innermost regions of the system. The dust
composition may thus vary as a function of the distance to the star,
and this explains why the scattering and emission cross sections
depend on $r$ in the previous equation.

\subsubsection{Spatial distribution and filtering}

Because of the limited constraints on the spatial distribution of the
dust, a 2D geometry is usually sufficient. If the disc is in addition
assumed to be axisymmetrical, its geometry can be fully described by a
1D radial distribution. Its appearance on the sky plane is then
defined by an inclination angle with respect to line of sight and a
position angle. The surface density profile can for example follow a
radial power law ($\Sigma(r) \propto r^{\alpha}$) or a smooth
combination of radial power laws. It can also come from
dynamical/collisional simulations. In most cases, the inclination of
the exozodi cannot be derived from the observations directly. It is
taken to be the same as that of the cold debris disc, if any, and
resolved.

As indicated in Sec.~\ref{nir}, the field-of-view of the near-IR
interferometric instruments (FLUOR at CHARA, PIONIER at VLTI) is
limited to a few au, which means that a fraction of the exozodiacal
emission can be filtered out. A synthetised flux $\Phi$ at a given
wavelength is then computed taking into account the telescope
transmission profile $T$ that falls to zero outside the FOV:
\begin{equation}
\Phi(\lambda) = \int_0^\infty 2 \pi \overline{T}(r) \Phi(\lambda,r)
\Sigma(r) r \mathrm{d}r,
\label{eq:transmission}
\end{equation}
where $\overline{T}(r)$ is obtained by first projecting the telescope
transmission map $T$ on the exozodi plane, and by then
azimuthally averaging along circles of radius $r$ in the disc frame
\citep{2013A&A...555A.146L}. This approach is valid as long as the
transmission profile is axisymmetrical in the sky plane, which is the
case for the FLUOR and PIONIER instruments. 

The sky transmission maps of the KIN and LBTI mid-IR nulling
experiments are, on the other hand, much more complex in shape, and
depend sensibly on observing parameters such as the observing
date/time, and the parallactic and hour angles
\citep[e.g.][]{2013ApJ...763..119M, Defrere:2015}. The above 1D
approach (Eq.~\ref{eq:transmission}) is therefore not appropriate, and
it is necessary to multiply the synthetic 2D disc image in the sky
plane with the 2D transmission map to obtain a synthetic null value.

\subsubsection{Additional refinements}

In some cases, additional refinements are required to improve the
modelling approach. For example, Vega is a star viewed nearly pole-on
known to be a rapid rotator. As a consequence, the star is wider and
cooler at the equator where the dust disc lies, and hotter at the
poles, with an effective temperature difference of about 2250\,K
\citep{2006ApJ...645..664A}. Therefore, the dust grains are
significantly less illuminated by the star's light than what one could
assume based on the pole-on, observed star spectrum. This has some
impacts on the grain temperatures and dust location in particular, and
this was accounted for in the models of \citet{2006A&A...452..237A}
and \citet{defrere2011}.

When it comes close to the sublimation distance, the competition of
the sublimation with other important physical processes (radiation
pressure, Poynting-Robertson drag, collisions) affects the dust
distribution. For example, micron-sized and smaller grains can suffer
a strong radiation pressure force around a bright star like Fomalhaut,
making the dust lifetime through this process smaller than the time it
takes to sublimate. To account for this effect,
\citet{2013A&A...555A.146L} assessed the sublimation temperature for
each grain size such that the sublimation timescale is equal to the
shortest dynamical timescale. The sublimation temperature is thus not
anymore a constant value but becomes size-dependent. This approach is
interesting because it introduces dynamical constraints into the
radiative transfer modelling, with a direct impact on the description
of the disc inner rim.

Finally, the limited number of observational constraints complicates
the overall modelling approach. It is also acknowledged that fitting a
spectral energy distribution is degenerate but one can work out
probabilities for each configuration using a Bayesian inference
method. A goodness of fit (usually a $\chi^2$) is computed for each
set of parameters and transformed into a probability assuming a
Gaussian likelihood function ($\propto e^{-\chi^2/2}$) for Bayesian
analysis. One can then obtain marginalised probability distributions
for each free parameter by projection of these probabilities onto each
dimension of the parameter space. This allows to find the best match
to the (multi-wavelength) observations and to calculate uncertainties
for the dust radial profile, the composition of the grains, their size
distribution, in particular the minimum grain size in the belt, and
the dust mass.

\subsection{Examples of modelled systems}

The radiative transfer modelling approach described above has been
applied to several systems around famous A-type stars (Sects.~\ref{vega}
and \ref{fom}), solar-type stars (Sect.~\ref{Stype}). Recently, radiation 
transfer modelling with a slightly different approach was performed on a large sample
(Sect.~\ref{sampletype}) to check whether all the dust
properties obtained for specific systems still hold when investigating
a wider range of exozodis.

\subsubsection{Vega}\label{vega}

Vega was the first system detected with hot dust using CHARA/FLUOR
near-IR, K-band observations and modelled using the methodology
described in subsection \ref{radjc} \citep{2006A&A...452..237A}. The
model was revisited in \citet{defrere2011} to account for the
IOTA/IONIC, interferometric H-band measurement, confirming the main
conclusions. It is found that the grains composing this exozodi are
submicron-sized, highly refractive (graphite or amorphous carbons) and
their size distribution index $\kappa$ seems steeper than the
usual -3.5. The grains are mainly located well within one au of Vega
and heated up to the assumed sublimation temperature of carbonaceous
grains (1900\,K). In fact, the emitting region locates typically
between 0.1 and 0.3\,au, consistent with the Palomar Fiber Nuller data
which suggest that any emission contributing to at least 1\% of the
near-infrared flux can arise only from within 0.2au
\citep{2011ApJ...736...14M}. Therefore, the grains appear to
accumulate near to their sublimation distance, with a surface density
profile that falls off rapidly with distance to the star, in sharp
constrast with the shallow density profile of the zodiacal cloud in
the Solar System (see subsection \ref{radialzd}). The dust mass is estimated to lie between a few
$10^{-9}$M$_\oplus$ and a few $10^{-8}$M$_\oplus$. The grains are
small and thus essentially unbound due to radiation pressure. The disc
is also dense enough for collisions to be frequent, with a timescale
of about a year. Therefore, Vega's hot exozodi needs to be replenished
at a rate of the order of $10^{-9}$M$_\oplus$/yr. The lack of
significant mid-IR excess below $\lambda \sim 15\,\mu$m, using either
spectro-photometry (Spitzer/IRS) or interferometry (BLINC/MMT, KIN),
does not give hints into the presence of a warm exozodiacal dust belt
at a few au from the star. Nevertheless, \citet{2013ApJ...763..118S}
argue that Vega could possess a $\sim 170\,$K belt at about 14au.

\subsubsection{Fomalhaut}\label{fom}

\begin{figure}
\centering
\includegraphics[scale=0.42]{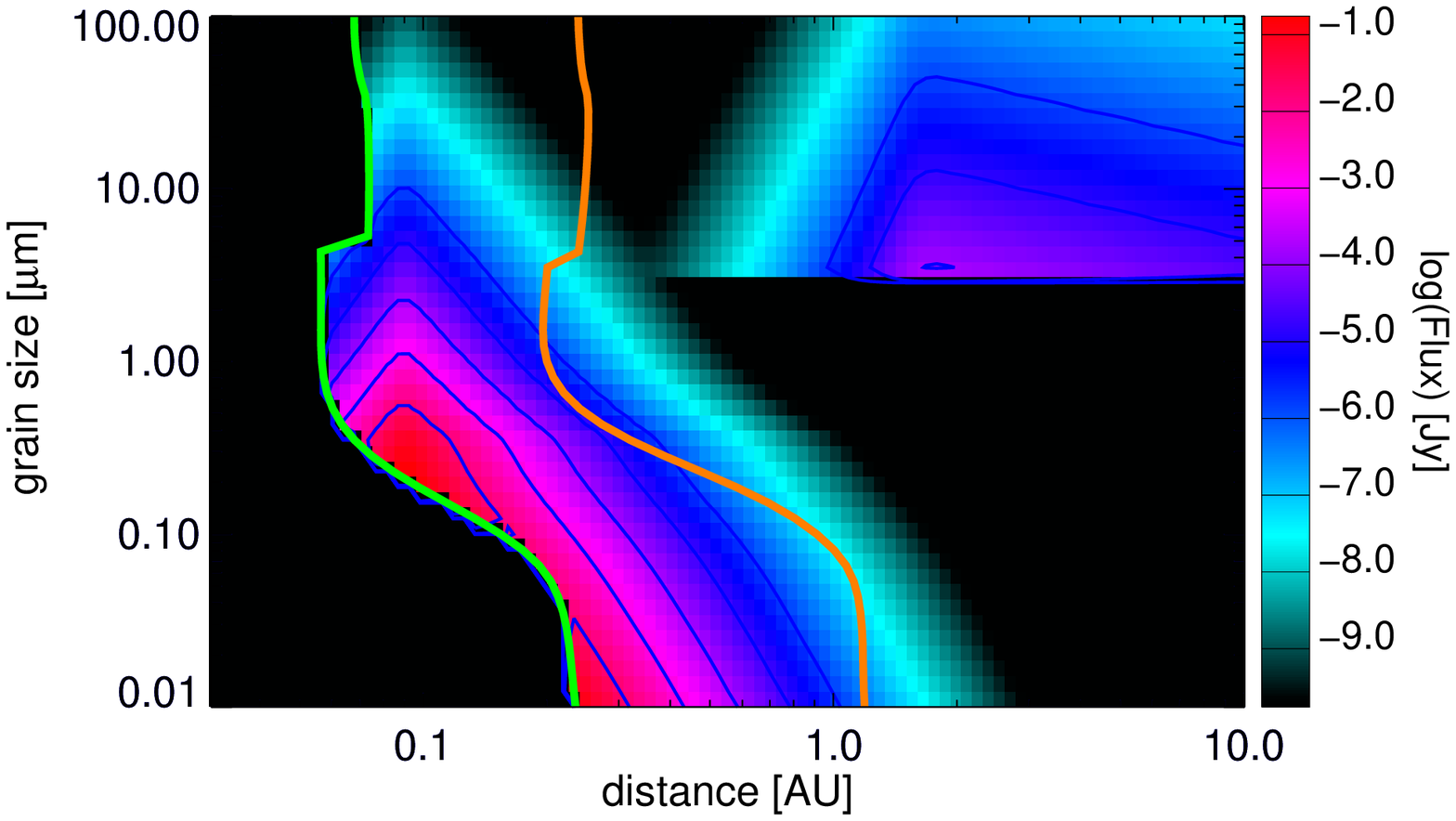}
\includegraphics[scale=0.42]{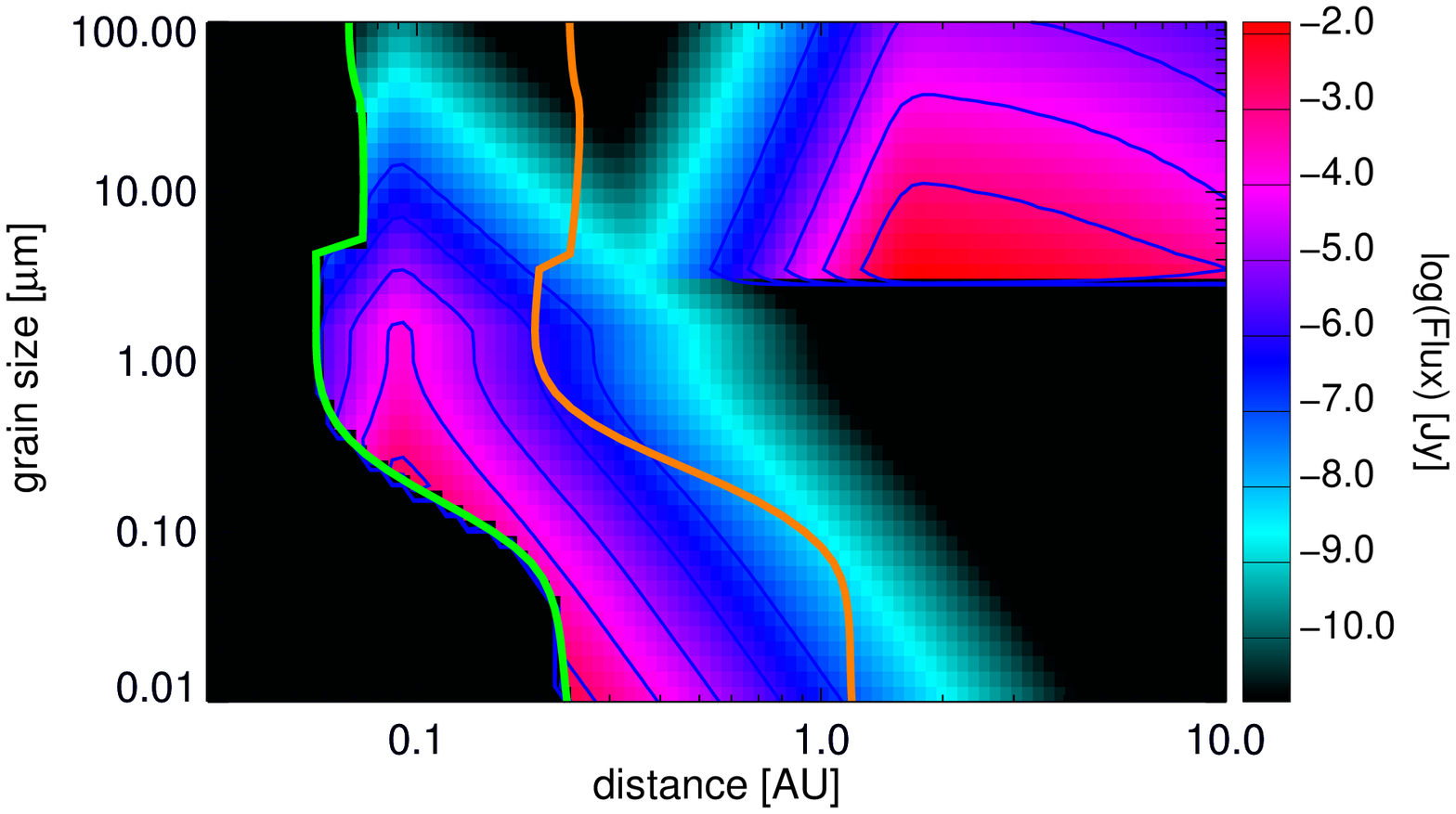}
\caption{Maps of the best fit model of the Fomalhaut exozodi showing
  the distribution of flux as a function of distance to the star and
  grain size. {\it Top}: $\lambda=2.18\,\mu$m, {\it bottom}:
  $\lambda=12\,\mu$m. The orange and green lines are showing the
  sublimation distance for silicate and carbon grains, respectively
  \citep{2013A&A...555A.146L}.}
\label{fig:radtra1}
\end{figure}

\begin{figure}
\centering
\includegraphics[scale=0.42]{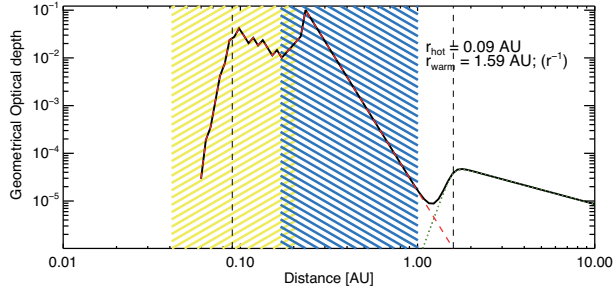}
\caption{Geometrical optical depth as a function of orbital distance
  for the best fit model found for the warm (in green) and hot (in red)
  populations around Fomalhaut \citep{2013A&A...555A.146L}. The dashed
  regions are for the sublimation zones of carbon (yellow) and
  silicates (blue).}
\label{fig:radtra2}
\end{figure}

A similar modelling approach was employed for Fomalhaut in
\citet{2013A&A...555A.146L}, with similar conclusions for the hot
exozodiacal dust component primarily probed by the VLTI/VINCI near-IR
instrument. The hot belt ($\sim2000$\,K) is dominated by very small
submicron-sized, carbon-rich grains at the carbon sublimation rim
($\sim$0.1--0.3au), and with a steep grain size distribution. Its
mass would be a few $10^{-10}$M$_\oplus$, with a replenishement rate
of almost $10^{-7}$M$_\oplus$/year. Interestingly, the exozodiacal
disc is also interferometrically detected in the mid-IR with the KIN,
and \citet{2013A&A...555A.146L} showed that a significant fraction of
mid-IR emission comes from a distinct belt of warm dust ($\sim
400$\,K) at $\sim$~2au dominated by micrometer-sized grains close to
the blow-out size, as can be seen in the bottom panel of
Fig.~\ref{fig:radtra1}. The subsequent radial dependence of the
optical depth is shown in Fig.~\ref{fig:radtra2}. The total dust mass
in the warm belt is estimated to a few $10^{-6}$M$_\oplus$, and it was
shown to be a possible source of carbonaceous dust grains for the hot
belt. However, the position and properties of the warm belt are
debated. Indeed, \citet{2013ApJ...763..118S,2016ApJ...818...45S}
suggest that the warm belt is further away, close to the ice
sublimation line at about 10\,au, which would make this belt
consistent with collisional evolution of an in situ planetesimal belt,
contrary to the scenario with a belt at 2\,au.

\subsubsection{Solar-type stars: $\tau$~Ceti and $\eta$~Crv}\label{Stype}

$\tau$ Ceti is one of the closest sun-like stars ($\sim$ 3.65pc) and
is known to host hot dust (CHARA/FLUOR, K-band excess), but does not
show any warm emission in the mid-IR \citep{2007A&A...475..243D}. The
best fit found for this system favours submicron-sized
grains at a few stellar radii (similar to the F-corona distance, see
subsection \ref{inzoobs}). The typical dust mass extracted from these
models is similar to Vega, i.e. on the order of $10^{-9}$M$_\oplus$.

\begin{figure}
\centering
\includegraphics[scale=0.42]{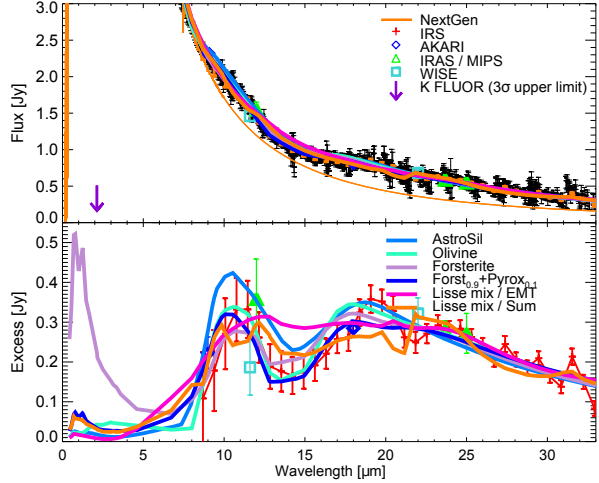}
\caption{Best-fitting exozodi models of the $\eta$ Corvi's SED for different grain compositions in the mid-IR.}
\label{fig:radtra3}
\end{figure}

\begin{figure}
\centering
\includegraphics[scale=0.52]{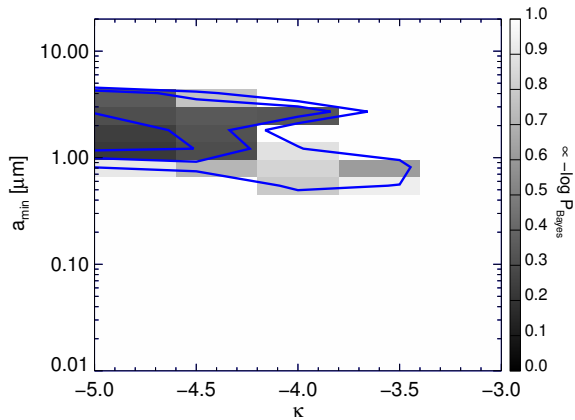}
\caption{Map of Bayesian probabilities for $\eta$ Corvi showing the
  most probable models in terms of minimum grain size versus slope
  of the size distribution $\kappa$.  The logarithmic display showing
  probabilities stretches between 0 and 1 and the smallest values are
  the best fit models.}
\label{fig:radtra4}
\end{figure}

The same approach was used to fit the $\eta$ Corvi warm belt using
nulls from the KIN and LBTI, Spitzer/IRS spectrum and the broadband
photometry in the mid-IR (WISE, AKARI, MIPS) \citep[][there is no detection in the K-band with
CHARA/FLUOR]{2016ApJ...817..165L}. The best-fit found (see
Fig.~\ref{fig:radtra3}) is a belt
located at $\sim 0.2$au made up of fosterite (high-albedo) grains
larger than $1\,\mu$m close to the blow-out size. Evidence for a size distribution that is steeper than predicted from collisional
equilibrium (see Fig.~\ref{fig:radtra4}) is puzzling as it results in an overabundance
of small grains. The total dust mass derived is about an order of
magnitude greater than for Fomalhaut's warm dust ($\sim 2\times
10^{-5}$M$_\oplus$).

\subsubsection{Modelling on a large sample}\label{sampletype}

\begin{figure}
\centering
\includegraphics[scale=0.4]{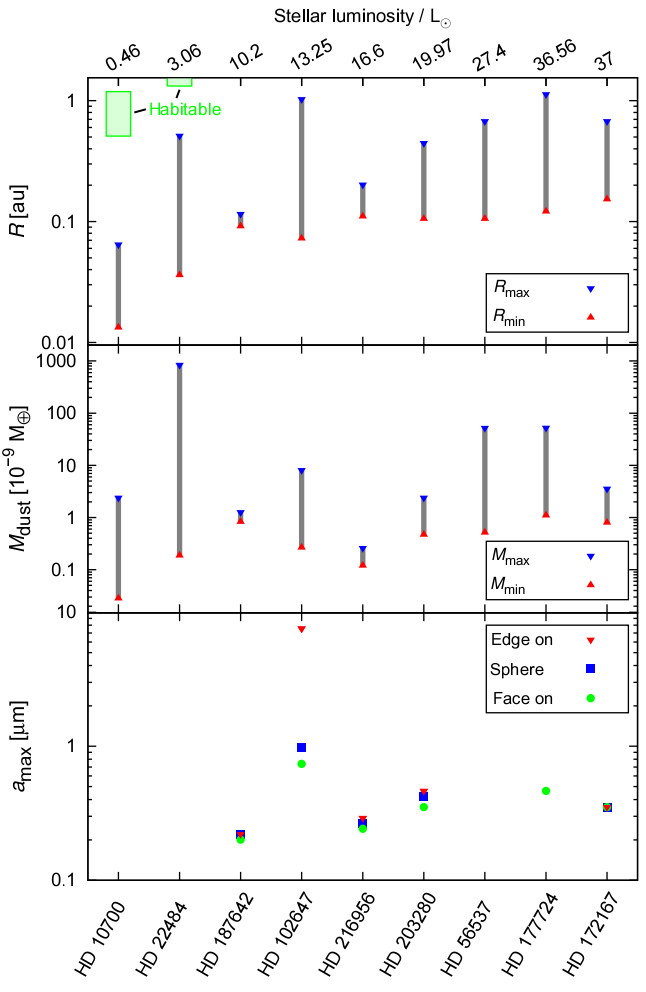}
\caption{Results of the fitting procedure of 9 exozodis \citep{Kirch17}. The stellar luminosity of these systems increases from left to right. {\it Top:} Minimum and maximum radius within which the exozodi is located (the HZ is shown as a green box when it fits in the frame). 
{\it Middle:} Minimum and maximum dust mass of the exozodis. {\it Bottom:} Maximum grain size (when it can be estimated from data) for different spatial configurations (edge-on, spherical shell and face-on).}
\label{fig:radtra5}
\end{figure}

A recent study by \citet{Kirch17} used an extended sample
of hot exozodis ($\sim$ 20 systems from H and K-band combined sample
described in subsection \ref{nir}) and radiative transfer modelling to put
more general constraints on dust properties and its location in these
systems. They also included published data at longer wavelengths in
the mid-IR and far-IR to better constrain these systems with hot
dust. The dust grain sizes and locations inferred from this large
sample are pretty similar to those found for the hot dust around
Fomalhaut (see \ref{fom}). For nine systems with the
tightest observational constraints, they conclude that the grain
sizes must be below 0.2-0.5 microns but above 0.02-0.15 microns and
that the hot dust must be located within $\sim$ 0.01 and 1au from the
star depending on its luminosity. Interestingly, they identify a
significant trend that dust distance increases with stellar
luminosity. The trend is such that all exozodis appear to have
approximately the same temperature. If confirmed, this might imply
that the hot dust phenomenon might be driven by temperature-driven
processes. Dust masses are estimated to be in the range (0.2-3.5)
$\times$ $10^{-9}$ M$_\oplus$ for this large sample. These results are summarised 
in Fig.~\ref{fig:radtra5}, where the names of these
systems can be found. Next, they demonstrate that a silicate
composition for the hot dust would be inconsistent with the observed
mid-IR fluxes and favour more absorbing materials (e.g. carbonaceous
material such as graphite) such as found by previous studies (see \ref{vega}, \ref{fom} and \ref{Stype}). 
They also show that the near-IR excesses
should be dominated by thermal radiation (though a contribution of
scattered light of up to 35\% cannot be excluded). Finally, they
predict that the polarisation degree of these discs should
always be below 5\%, which agrees with recent polarisation
observations (see subsection \ref{nir}).

\section{Origin of exozodis}\label{origin}

There are schematically two strong observationally-derived constraints on the origin of exozodis, one regarding their collisional lifetime and another one related to the size of the observed dust.

The first constraint is that the amount of dust that is observed in most exozodi-hosting systems should rapidly collisionally erode and be ultimately blown out by radiation pressure. This dust has, thus, to be replenished on relatively short timescales. An obvious mechanism would be an in situ collisional
cascade starting from much larger, and undetectable, parent bodies that would act as a mass reservoir for the system. This scenario is the canonical explanation for the evolution of ``classical'' cold debris discs, which in steady state can usually sustain the observed level of dustiness for the age of the
system \citep[e.g.,][]{2008ApJ...673.1123L}. However, for exozodis, because the dust is located in regions where collision rates and collision velocities are expected to be much higher, and thus the collisional erosion of a parent body belt much faster, the steady collisional cascade scenario
encounters some problems. This scenario is indeed only viable for warm exozodis around young ($<120$Myr) stars, for which the collisional lifetime of planetesimal-like parent bodies can become close to the age of the system. It fails to explain the few cases of warm exozodis 
around older stars \citep{2013MNRAS.433.2334K}, and, more crucially, it cannot explain the dustiness of hot exozodis, for which the observed near-IR
excess levels cannot be sustained by steady state in situ collisional cascades regardless of the age of the system \citep{2007ApJ...658..569W, 2013A&A...555A.146L}.

Another important constraint is that, amongst the exozodis for which a detailled fit and modelling of the grain population has been undertaken so far, the majority have turned out to harbour a vast population of tiny submicron grains (see section \ref{grater}). For instance, for Vega and Fomalhaut, 
the submicron grains are found in hot exozodi belts very close to their host stars (see subsection \ref{vega} and \ref{fom}), and for HD 172555 \citep{johnson2012}, the tiny grains are located in a warm belt 
further out at a few au\footnote{We note that for $\eta$ corvi, submicron grains might also be present, even though the best grain population fit does not require their presence (see subsection \ref{Stype} and Fig.~\ref{fig:radtra4}).}. These sub-$\mu$m populations are very difficult to explain given the fact that such 
grains should be blown out by radiation pressure on dynamical timescales, i.e., less than a few years at the radial distance they are observed. Their presence is, therefore, a challenge to classical dust evolution scenarios.

\begin{figure}
\centering
\includegraphics[scale=0.35]{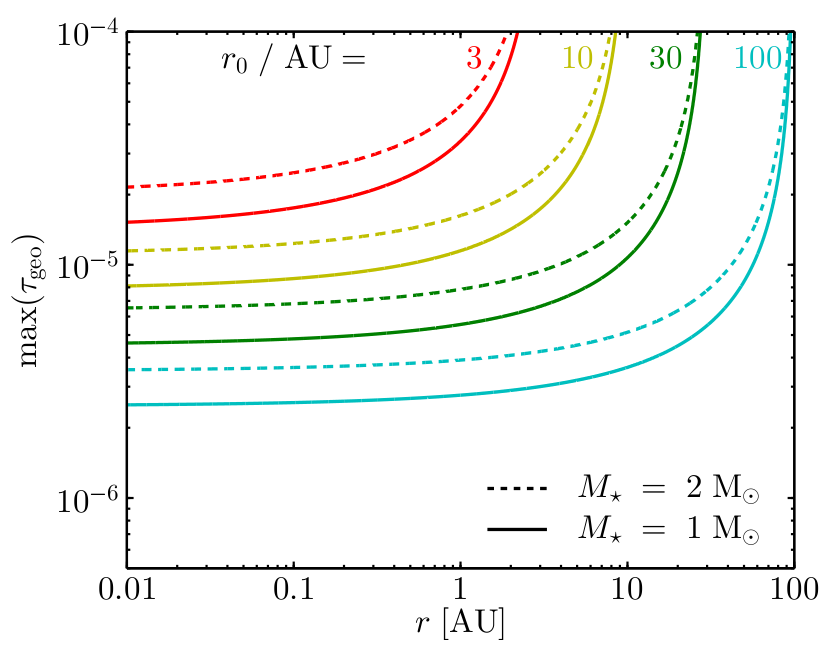}
\caption{Maximum geometrical optical depth (proportional to observed fractional luminosity) created from grains migrating inwards under the effect of PR-drag as a function of distance to the star \citep{2014A&A...571A..51V}. The parent belt from which grains start migrating is located at different $r_0$ (which corresponds to different colours)
for two different stellar types (solid and dashed lines).}
\label{fig:prdrag}
\end{figure}

\subsection{PR-drag}\label{prdr}
Grains migrating inwards by PR-drag from an outer cold belt are the most obvious way to feed in an inner belt continuously from a large reservoir of mass (it feeds the inner region of our zodiacal cloud, see Fig.~\ref{fig:zodi_sketch}). However, PR-drag on its own is not efficient enough to provide a sufficient amount of
hot dust that could explain near-IR observations \citep[e.g.][]{2006A&A...452..237A,2014A&A...571A..51V}.

\begin{figure*}
\centering
\includegraphics[scale=0.4]{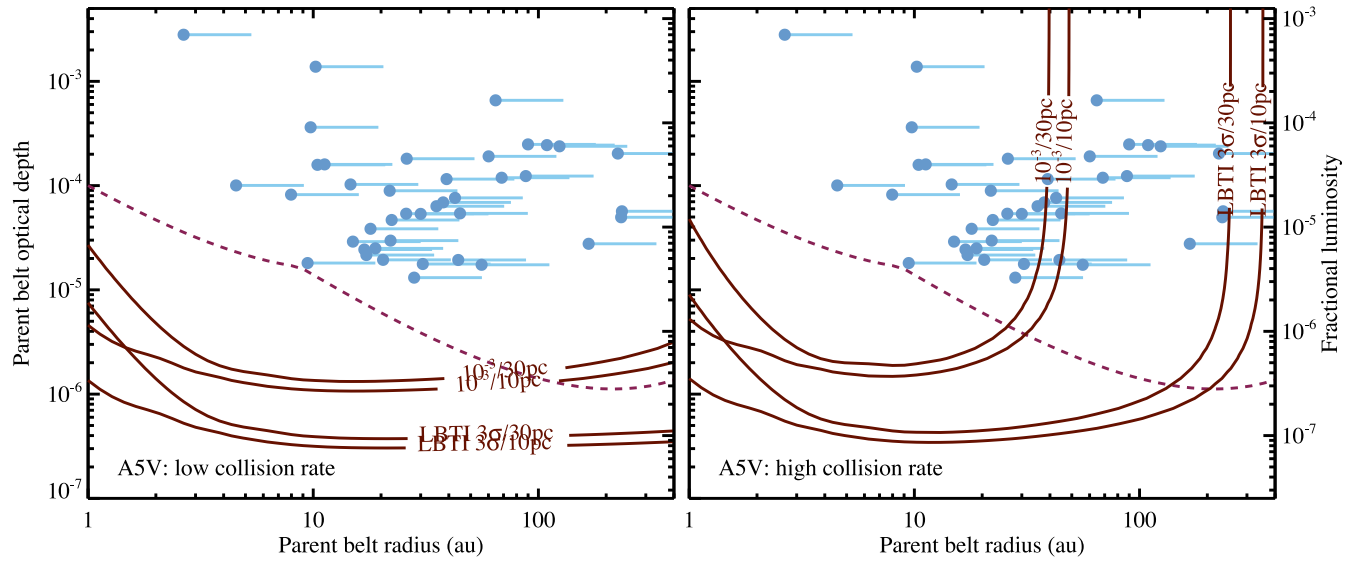}
\caption{LBTI null excesses for a range of parent belt parameters around A stars, for two different sensitivities $3\times 10^{−4}$ (LBTI-like) and $10^{−3}$, and two stellar distances (10 and
30pc, contours as labelled). The left panel assumes a low collision rate in the parent belt and the right panel a higher collision rate \citep[see][for details]{2015MNRAS.449.2304K}. 
The blue dots show known cold debris discs and connected lines show the probable increase in size due to non-blackbody emission from small dust. Dashed lines show detection limits for debris discs around nearby stars. 
The expected null excesses always increase with parent belt optical depth, so for each distance the contours show the minimum parent belt optical depth
for warm dust to be detectable at that level \citep{2015MNRAS.449.2304K}.}
\label{fig:prdraggrant}
\end{figure*}

On the other hand, these grains migrating by PR-drag will naturally lead to some level of warm dust that may be detectable in the mid-IR. As explained in subsection \ref{sec:mir}, mid-IR excessess are correlated with the presence of an outer reservoir and,
from the KIN survey, the detected mid-IR levels tend to be similar \citep[see Table 2 in][]{Mennesson:2014}. This favours PR-drag as a good explanation as it is relatively insensitive to the properties of the parent belt. In addition, \citet{Mennesson:2014}
show that the observed mid-IR excessess are compatible with PR-drag levels expected from \citet{2005A&A...433.1007W}.

Fig.~\ref{fig:prdrag} shows the vertical optical depth as a function of the radial distance to the star that is expected from the interplay between PR-drag and collisions from an outer belt located at different initial positions (different colours) and for two different stellar masses (dashed and solid lines).
The maximum optical depth reached in inner regions is always close to $\sim 10^{-5}$ (within a factor 3), which is in agreement with the KIN observations showing a roughly constant mid-IR level for different systems. 
\citet{2015MNRAS.449.2304K} showed that the LBTI will be sensitive to this PR-drag produced dust. Indeed, as shown in Fig.~\ref{fig:prdraggrant}, the level of warm dust produced by PR-drag from debris discs that are detected today (blue dots) is detectable with the LBTI (solid lines).
Note that the PR-drag warm dust produced from a parent belt with a small $10^{-6}$ optical depth could even be detected with the LBTI while the actual cold disc cannot currently be detected with photometry (dashed line).    
Dust levels in inner regions of planetary systems can be hundreds of times higher than in our Solar System and cannot be considered as insignificant. This amount of dust is detrimental to future missions trying to discover new Earth-like planets
(see subsection \ref{hamper}). However, the presence of a planet in the system
may prevent dust from reaching the habitable zone. Thus, a non-detection of this warm dust for
debris disc systems observed with the LBTI may well be the indirect evidence of a planet sitting in between the HZ and the cold belt \citep{2015MNRAS.449.2304K}. 

\subsection{Pile-up of sublimating dust}\label{pileupsub}

PR-drag causes dust to migrate inwards from a parent belt where it is produced (see subsection \ref{prdr}).
However, the amount of dust that reaches the inner parts of a planetary system in this way
is limited by destructive mutual collisions between the migrating dust grains \citep{2005A&A...433.1007W}.
While the population of dust set by the balance between PR-drag and collisions is insufficient to explain hot exozodiacal dust observations, this conclusion may be averted
if a mechanism exists that prolongs the residence time of dust close to the star, piling it up above the normal PR-drag population.

\begin{figure}
\centering
\includegraphics[scale=0.3]{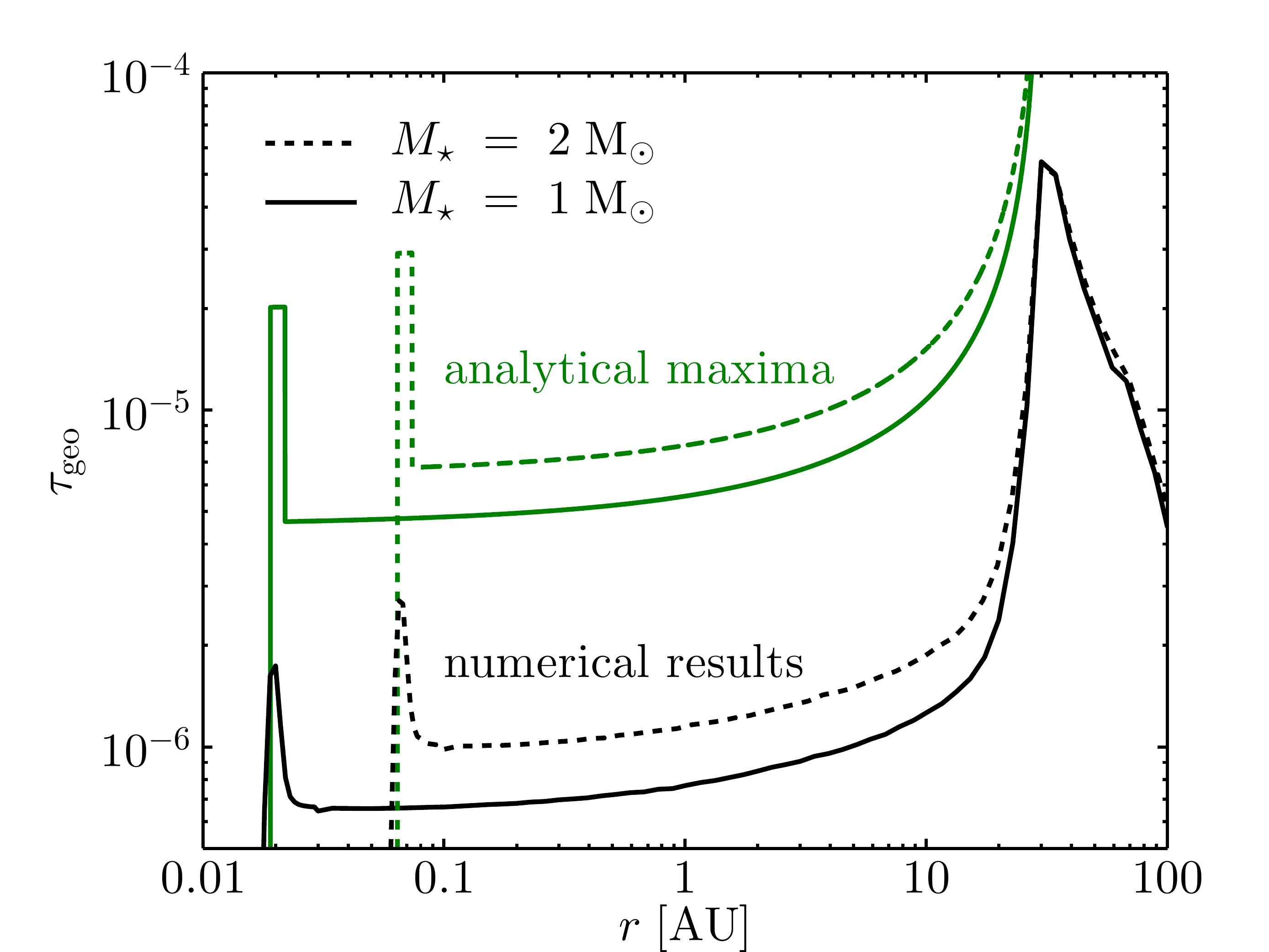}
\caption{Radial profiles of geometrical optical depth for discs undergoing evolution due to collisions, PR-drag, and sublimation. Parent belts are located at 30au; PR drag tails extend inwards. The pile-up of dust due to sublimation causes the enhancements at small radii \citep{2014A&A...571A..51V}.}
\label{fig:num_tau}
\end{figure}

One possible pile-up mechanism is the interplay of dust sublimation and radiation pressure forces, as studied in detail by \citet{2008Icar..195..871K,2009Icar..201..395K,2011EP&S...63.1067K}. When PR-drag brings dust grains to within a few stellar radii of their host star, 
significant sublimation sets in, decreasing the size of the dust grains. As a result, radiation pressure becomes more important for these grains (with respect to gravity) and consequently their orbital eccentricities and semi-major axes increase. 
This slows down the inward migration of the dust due to PR-drag, with the net effect of creating a ring of piled-up dust around the sublimation radius (see Fig.~\ref{fig:num_tau}) as observed in the zodiacal dust at 4 solar radii (see subsection \ref{radialzd}).
However, the pile-up is only a factor of a few in terms of dust cross-section \citep{2009Icar..201..395K}.
Hence, this mechanism makes little difference to the observable signatures of dust in the sublimation zone and cannot explain hot exozodis \citep{2014A&A...571A..51V}.

\subsection{Trapping with gas}

Fomalhaut multiwavelength observations are best explained with two belts in its inner region, a hot belt at a fraction of au (composed of very small unbound grains close to the sublimation rim) and a warmer belt at $\sim$ 2au (see subsection \ref{grater}). \citet{2013A&A...555A.146L} suggest that 
the hotter population is due to a pile-up of dust owing to sublimation as explained in subsection \ref{pileupsub}. They find that they can explain the mass in the hot belt by assuming that grains migrate inwards by PR-drag from the warm belt. However, they cannot explain the high optical depth derived
from observations as they would need to keep unbound grains for longer than a dynamical timescale. They show that gas drag could work to slow down these unbound grains when assuming that gas is produced from sublimation of grains for the age of the system. The amount of gas
required is $\sim 5\times 10^{-3}$M$_\oplus$ and could also be coming from gas produced farther away in the cold belt and viscously evolving to create an accretion disc up to the star \citep{2016MNRAS.461..845K,2016MNRAS.461.1614K}.

Another possibility is that gas drag could act further out than the sublimation rim. Indeed, for most debris discs, an atomic gas disc is expected to go all the way to the star \citep{2017Kral} and may act to brake the grains coming in under the effect of PR-drag. The point where
the grains have a steady orbit then depends on the amount of gas in the system. This new potential scenario could then explain an enhancement of warm dust compared to PR-drag alone (Kral et al, in prep.).

\subsection{Dynamical Instabilities}
A planetary system that has recently undergone a dynamical instability could lead to high levels of dust production, in a similar manner to the Solar System's Late Heavy Bombardment. 
\citet{2009MNRAS.399..385B} show that a LHB leads to high levels of dust throughout the planetary system, and a particularly increased emission in the near/mid-IR. They use Spitzer 24$\mu$m observations to limit the occurence of LHB-type events to $<$ 4\%.
However, in general this dust is short lived. \citet{2013MNRAS.433.2938B}
used a suite of simulations in which planetary systems went unstable, to show that the high levels of dust generated in the inner regions of planetary systems are insufficiently long lived to explain the high incidence of exozodiacal dust observed, 
even if every star had an instability at some point during its lifetime. It remains possible that individual exozodis are produced following such a dynamical instability. 

\begin{figure*}
\includegraphics[width=0.99\textwidth]{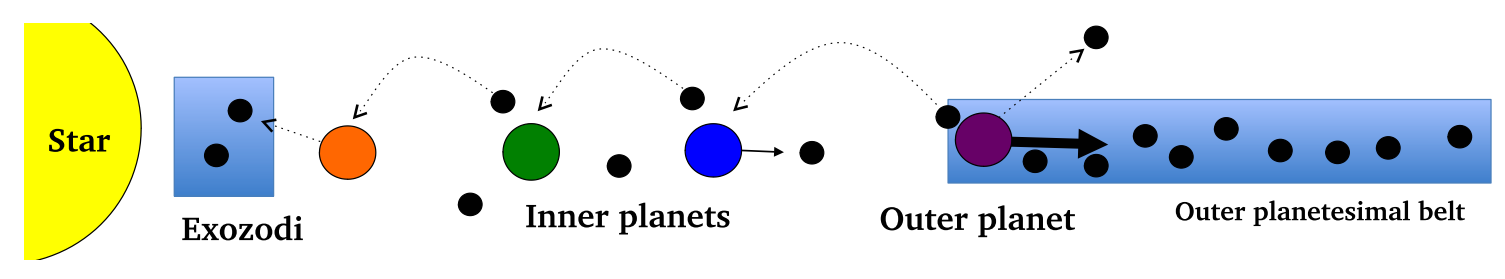}
\caption{Schematic of the process to create an exozodi from an outer cold belt through planetesimal driven migration \citep{2014MNRAS.441.2380B}. }

\label{fig:pdm1}
\end{figure*}

\subsection{Material scattered inwards by planets}
Large masses in planetesimals can survive in the outer regions of planetary systems, where collisional evolution timescales are longer. If this material is transported inwards, it could supply the observed high levels of dust in the inner regions of planetary systems. 
One transport mechanism is scattering by planets (Fig.~\ref{fig:pdm1} illustrates a cartoon schematic of the scenario). In order for sufficient material to be scattered inwards on long timescales the planets must be low mass ($<$ tens of earth masses) and the planetary system must be sufficiently full for scattering
 to occur \citep{2012A&A...548A.104B,2017MNRAS.464.3385W}. If the planet masses are higher, the scattering process is shorter lived, and therefore, unless the architecture of the planetary system has altered during its evolution, cannot supply the observed high levels
 of dust in old (Gyr) systems. The planetesimal belts most suited to supplying the observed exozodis are very massive and at large radii. In order for planetary systems to supply 
the observed exozodiacal dust, the constraints on the architecture of the planetary system required are very tight, and it, therefore, seems unlikely to be the case for a significant proportion of planetary systems.

\subsection{Planetesimal driven migration}
\begin{figure}
\centering
\includegraphics[width=0.5\textwidth]{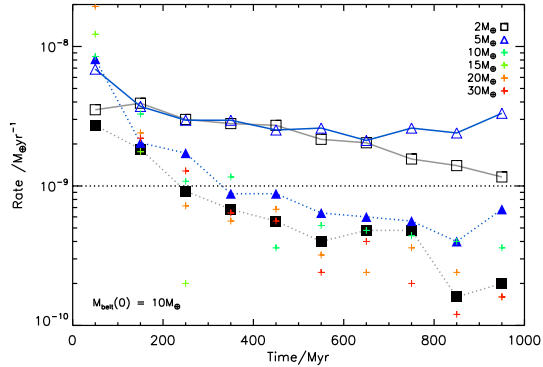}
\caption{Planetesimal scattering, following planetesimal driven migration is sufficiently efficient to sustain exozodis \citep{2014MNRAS.441.2380B}. The rate at which particles are scattered interior to 3au in a suite of simulations where the planet migrated outwards (open)
should be compared to the nominal rate required to sustain Vega's exozodi (horizontal dotted line) and the scattering rates without migration (closed).}
\label{fig:pdm2}
\end{figure}

For planets of comparable masses to planetesimal belts, scattering of planetesimals can drive migration, in either the inwards or the outwards direction. If the migration is sustained, the planet encounters fresh material that can be scattered. Under the right conditions this
 can enhance the supply of material scattered into the inner regions of a planetary system at late times \citep{2014MNRAS.441.2380B,2014MNRAS.442L..18R}. Planetesimal driven migration is stalled for planets above a critical mass, given approximately by  
Eq. 58 of \citet{2012ApJ...758...80O} or Eq.~1 of \citet{2014MNRAS.441.2380B}. The critical mass is proportional to the local surface density of the disc. For low mass planets (tens of earth masses) migration can be sustained on Gyr timescales. As the planet migrates out through the disc, it continues to scatter material inwards, potentially
to supply an exozodi, if further planets orbit interior to the belt (see Fig.~\ref{fig:pdm1}).  Fig.~\ref{fig:pdm2} indicates the rate at which material is scattered 
interior to 3au, in a simulation in which a 5M$_\oplus$ planet in a belt of initial total mass 10M$_\oplus$, migrates from 15au to 25au. Four interior planets of 5M$_\oplus$ scatter particles interior to 3au. The open shapes indicate simulations where the outer planet migrated, whilst the filled 
shapes and crosses indicate simulations in which the planet did not migrate as the planetesimals were considered to be test particles. The horizontal dotted line indicates the approximate scattering rate required to sustain exozodiacal dust estimated for the Vega system of $10^{-9}$M$_\oplus$yr$^{-1}$. 
This simulation shows that planetesimal driven migration has the potential to supply exozodiacal dust. 

\subsection{Inner MMRs of eccentric planets}

\begin{figure}
\centering
\includegraphics[scale=0.42]{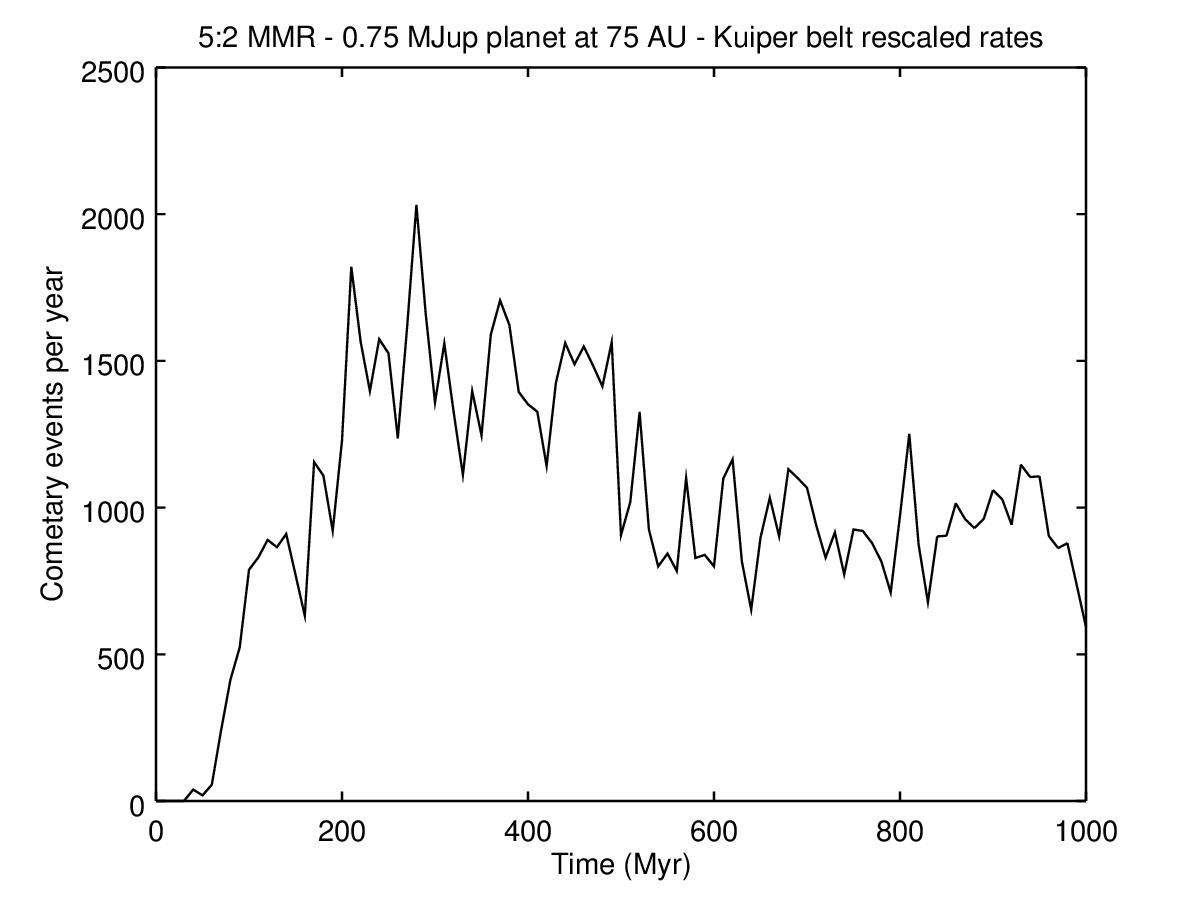}
\caption{Comet production rate calculated via a numerical simulation of a Kuiper Belt analogue with a 0.75-Jupiter-mass planet located at 75au, showing that the production of comets can be sustained for Gyr-timescales \citep{2016arXiv161102196F}.}
\label{fig:far16_fig8}
\end{figure}

Another mechanism to deliver comets into the inner regions of a planetary system
was recently proposed by \citet{2016arXiv161102196F}.
The mechanism involves a planet on a moderately eccentric orbit ($e_\mathrm{p}\gtrsim0.1$), located exterior to a planetesimal belt.
If a mean-motion resonance (MMR) of the planet overlaps with the belt,
the orbital eccentricities of particles trapped in resonance can be pumped up \citep{1996Icar..120..358B}.
The maximum eccentricity that can be reached in this way depends on the resonance.
In some cases (e.g., the 4:1 MMR), the mechanism can place planetesimals on cometary orbits directly ($e\sim1$).
In other cases (e.g., the 5:2 MMR with a distant outer planet), comets may not be produced directly, but (depending on the mass of the planet)
the eccentricities can be pumped up to sufficiently high levels that the planetesimals reach the chaotic zone of the planet.
When this happens, the planet-crossing bodies have some probability of being scattered on cometary orbits.
Given the long timescale of the eccentricity pumping and the relatively low probability of scattering,
this process can generate comets over very long ($\sim$Gyr) timescales (see Fig.~\ref{fig:far16_fig8}).
Hence, it provides a possible explanation for exozodis occurring around older stars as well as younger ones.
Furthermore, the mass input rate of comets, generated from a Kuiper Belt analogue
and sustained over these long timescales, is estimated at $\sim$10$^{-12}$ to 10$^{-11}$M$_\oplus$/yr,
more than enough to sustain a population of dust comparable to the Solar System's zodiacal cloud.
In this mechanism, the planet should be located outside of the belt at 10's of au, which is less likely than a closer planet. However, double belt systems such as in our Solar System are rather common \citep{2009ApJ...699.1067M,2014MNRAS.444.3164K}. The planet
feeding the hot dust could be in interaction with the innermost belt as it is more likely to have planets closer in. 

\subsection{Trapping with MMRs}
Grains migrating in under PR-drag may be trapped on their way in mean-motion resonances with planets. This produces clumpy circumstellar rings of dust such as observed around the Earth \citep{dermott-et-al-1994,1995Natur.374..521R,2010Icar..209..848R} 
or Venus \citep{2007A&A...472..335L,2013Sci...342..960J} in our Solar System (as already mentioned in subsection \ref{radialzd}). This resonantly trapped dust should exist in extrasolar systems and may be detectable with repeated LBTI observations \citep{2015MNRAS.448..684S}.
The resulting dust levels can be higher than PR-drag levels as the dust is trapped for longer than the PR-drag timescale. However, as discussed in \citet{2015MNRAS.448..684S}, the libration timescale of trapped dust is still short and resonantly captured rings
can only account for a brightness excess of no more than a factor 10 compared to the background disc.

\subsection{Massive collisions}
\citet{2007ApJ...658..569W} showed that for a belt of a given age, there is a maximum possible disc mass that can be reached, since more massive discs will process their mass faster \citep[see also][]{2008ApJ...673.1123L}. However, some detections in the mid-IR are well above that
maximum mass (by a factor $>$ 1000), making it impossible to explain these large excesses with a steady state collisional cascade. \citet{2007ApJ...658..569W} deduce that most of these systems are likely undergoing some transient events.

Massive collisions such as the Moon forming collision with Earth are expected in the late stages of planetary formation \citep[e.g.][]{2006AJ....131.1837K,2009Icar..203..644R} and can potentially explain some high mid-IR excesses \citep{2012MNRAS.425..657J}. Assuming that after the impact, 30\% of the mass is in
mm-cm sized vapour condensates and 70\% in large planetesimals up to 500km, \citet{2012MNRAS.425..657J} find that the condensates deplete collisionally in about 1000yr, while the bigger boulders collide to create dust
that could be observed from other stars for $\sim$25Myr (with an equivalent of Spitzer at 24$\mu$m) after the impact at emission levels comparable to some observed warm dust systems. 

\begin{figure*}
\centering
\includegraphics[scale=0.35]{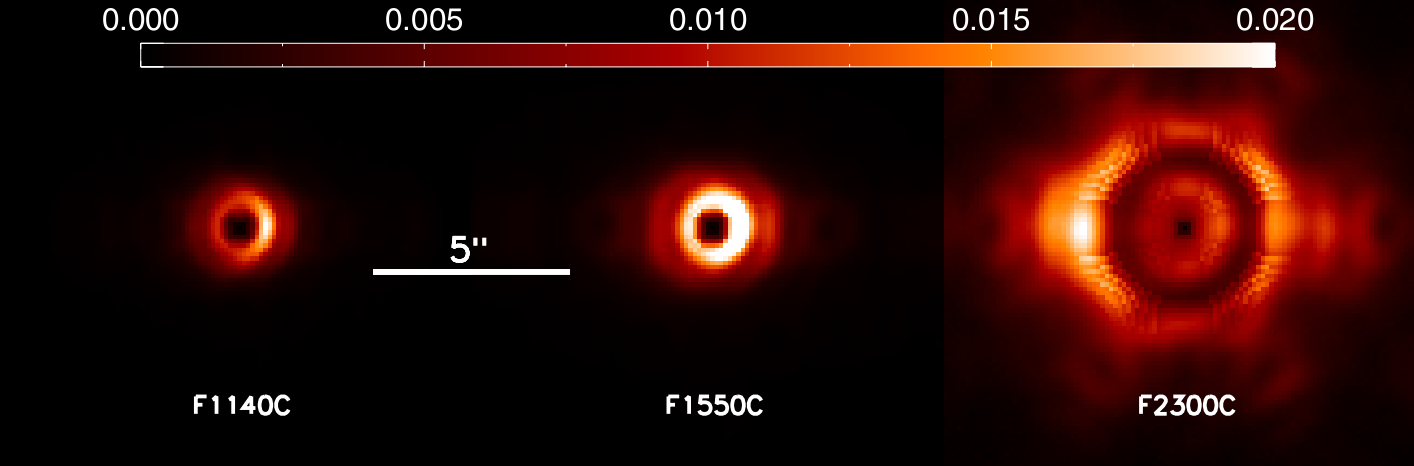}
\caption{MIRI/JWST synthetic observation of the aftermath of a massive collision between two asteroid-like objects. The
brightest part of the disc on the right side (collision point) at 11 and 15 microns and on the left side at 23 microns is a typical signature of these events \citep{2015A&A...573A..39K}.}
\label{fig:giantim}
\end{figure*}

\citet{2015A&A...573A..39K} also showed that these type of giant impacts can produce debris for millions of years that could be detectable with JWST/MIRI (see a synthetic image of such an observation in Fig.~\ref{fig:giantim}). These giant impacts are predicted to have a specific signature, which
is easily seen in Fig.~\ref{fig:giantim}. At the impact point there is an overdensity, which translate as a larger thermal emission in the mid-IR. For the specific case of MIRI at 23$\mu$m, due to the 2'' Lyot coronagraph set up at that wavelength, the inner region where the impact happens are not
observable and the asymmetry moves to the other side of the impact point (right subplot in Fig.~\ref{fig:giantim}) in the dust halo component. Detecting this specific signature would be a way to detect on-going planetary formation.

This type of massive collisions cannot explain the biggest mid-IR excesses \citep{2007ApJ...658..569W} but could account for a few. The probability of having a giant impact in a given system at a given time is too small to explain the detection rate of these mid-IR excesses on its own.

\subsection{Disintegrating rocky planets}

Another possible source of dust at distances of a few stellar radii is disintegrating rocky planets.
So far, three such objects were discovered in the \textit{Kepler} data: KIC~12557548b, \mbox{KOI-2700b}, and \mbox{K2-22b} \citep{2012ApJ...752....1R,2014ApJ...784...40R,2015ApJ...812..112S}.
Their light curves display dips that are strictly periodic, but highly variable in depth and have asymmetric profiles.
These are interpreted as the transits of a cloud of dust emitted by a small (sub-Earth-sized) evaporating planet.
The planets orbit at just a few stellar radii from the star, where the intense stellar irradiation sublimates the solid surface of the planet.
This leads to a gaseous outflow in which dust grains condense out as the gas expands and cools \citep{2012ApJ...752....1R,2013MNRAS.433.2294P}.
The dust is carried away from the planet, in a comet-like tail, where it is gradually destroyed by sublimation.
Estimates of the typical grain sizes cluster around 0.1 to 1$\mu$m \citep[e.g.,][]{2012A&A...545L...5B,2013A&A...557A..72B,2015ApJ...800L..21B,2016arXiv160900275V}.
The dust mass production rates are estimated at $10^{-11}$ to $10^{-9}$\,M$_\oplus$yr$^{-1}$ \citep[e.g.,][]{2012ApJ...752....1R,2014ApJ...784...40R,2013ApJ...776L...6K,2014A&A...572A..76V,2015ApJ...812..112S}.

There are two main problems with disintegrating rocky planets as an explanation for (all) hot exozodis. 1)~These bodies are likely quite rare, with an occurence rate estimated at $\sim$10$^{-4}$ from \textit{Kepler} statistics \citep{2013MNRAS.433.2294P}. 
This is much less frequent than hot exozodis. 2)~Given transit depths up to $\sim$1\% and orbital distances of several stellar radii, the dust covering fraction (and hence the fractional luminosity) is at most $\sim$10$^{-4}$, much lower than most hot exozodi detections. 
Note that this assumes the comet-like tail seen in transit is the only hot dust producing near-IR excess.


\subsection{Trapping with magnetic fields}\label{magn}

The \{PR-drag + sublimation\} scenario described in subsection \ref{pileupsub}, despite proposing a tantalising way of transporting material towards the sublimation distance and producing a pile-up of submicron grains there, cannot produce a near-IR excess that is high enough to explain the $\sim0.01$ 
value observed for hot exozodis. The main  problem is that the residence time of sublimating grains in the inner ring is not long enough to create a pile-up of the required level to explain the near-IR flux \citep{2014A&A...571A..51V}. 

To overcome this caveat, \citet{Rieke(2016)} proposed to add one crucial ingredient to the dynamics of small grains in the innermost disc regions, that is stellar magnetic field (that seems to play a role in our zodiacal cloud, see subsection \ref{submzd}). Indeed, around hot A stars, submicron grains could rapidly become charged by the photoelectric effect and thus 
interact with a potential magnetic field through Lorentz forces. While such magnetic fields are often thought to be non-existent around early-type stars, this simplified view has been revised by the discovery of a weak field around Vega \citep{2009A&A...500L..41L}. \citet{Rieke(2016)} showed that, for a 
typical A star, the stellar magnetic field is expected to rotate with the star out to a distance that is close to the sublimation radius. Under such circumstances, they estimate that grains of sizes $s\lesssim 100$nm can become temporarily trapped, on epicyclic orbits, by Lorentz forces for fields as 
low as 0.1G, which is less than the strength measured for Vega. The crucial point is that this magnetic trapping can last much longer than the sublimation time and thus produce and maintain a much more pronounced pile-up at the sublimation radius. For a typical inward flux (by PR-drag) of 
material from an asteroid-like belt similar to that of \citet{2014A&A...571A..51V}, \citet{Rieke(2016)} find that this pile-up could be high enough to produce the observed near-IR excesses.

Using these results, \citet{2016ApJ...818...45S} built an empirical model to fit the inner 20au excess of the Fomalhaut debris disc with four components: an asteroid belt, a PR disc of $\sim \mu$m-sized grains flowing inward from this belt, a pile-up ring of $\sim \mu$m silicate grains similar to that 
of \citet{2014A&A...571A..51V} and, finally, a hot ring of magnetically trapped carbon nano-grains. Using simple analytical estimates, they showed that a hot-ring resident time of $\sim5-10$yr is needed to reproduce the K-band excess, which is compatible with, though slightly above, 
the values derived by \citet{Rieke(2016)}.

There remain, however, several issues that need to be addressed to assess the viability of the magnetic-trapping scenario. The main one is probably to estimate if grains arriving at the sublimation distance can become small enough ($s\lesssim100$nm) before being blown out by radiation 
pressure. Indeed, the \citet{Rieke(2016)} model implicitly assumes that grains are immediately transformed into $<100$nm particles when arriving at the sublimation limit, but this size is more than one order of magnitude less than the blow-out size by radiation pressure, which is the expected 
size for PR-drifting grains coming from an asteroid belt further out. There is thus a risk that, if grains spend too much time in this intermediate $0.1-1\mu$m size range, they will be ejected from the sublimation zone \emph{before} being able to be trapped by Lorentz forces. Another potential caveat is 
that collisional destruction within the magnetic-trapping ring could shorten the residence time of nanograins. Estimates by \citet{Rieke(2016)} show that, for a wide range of grain sizes, the collisional timescale might indeed be one order of magnitude lower than
the magnetic trapping timescale, which might reduce the pile-up density to values too low to explain the observed fluxes. 
Another potential problem is that the Lorentz force expression used
by \citet{Rieke(2016)} (their Eq. 18) apparently does not include the stellar
wind velocity term $-\vec{v}_{sw} \times \vec{B}$ \citep[e.g.][]{1979P&SS...27.1269M,krivov-et-al-1998b,2003ApJ...594.1049R,2010ApJ...714...89C,2016ApJ...828...10L}.
This term is likely to be small for the case they modelled, since the winds of A-stars are very weak. As applied to other stellar types, their analysis is appropriate only to purely radial 
stellar magnetic fields and it remains to be seen how the trapping effect is affected by more realistic magnetic field geometries.
These issues should be investigated with detailed numerical simulations.


\section{Prospects for the future}\label{futpro}

The immediate prospect is the HOSTS survey \citep[in the mid-IR,][]{2015ApJS..216...24W} that has been recently started with the LBTI (see Section~\ref{sec:mir}). The goal of HOSTS is not only to determine the faint end of the luminosity distribution function but also to know which individual stars have the least amount of zodiacal dust. 
The HOSTS survey will bring valuable information on the correlation between warm dust and other key parameters of planetary systems, such as age, presence of an outer belt, and spectral type of the star. 

Another urgent prospect is to understand the variability of some of the hot exozodis. Variability on timescales of at least as short as
one year has been suggested for some exozodis \citep{2016A&A...595A..44E}. Repeated short period measurements will obtain the periodicity of these variations and will eventually unravel the origin of this variability. 

Observationally, testing the proposed scenarios for the origin of exozodis in section \ref{origin} is also a priority. For instance, for the scenario of dust trapping with magnetic field presented in subsection \ref{magn}, one could try to characterise the magnetic fields of stars with and without hot dust detected. 
Some measurements were already obtained by \citet{2011PASA...28..323W,2014MNRAS.444.3517M} 
but the overlap with exozodi host stars is small. A dedicated observational program should be started to look for magnetic field related correlations with exozodis. 

\begin{figure}
\centering
\includegraphics[scale=0.55]{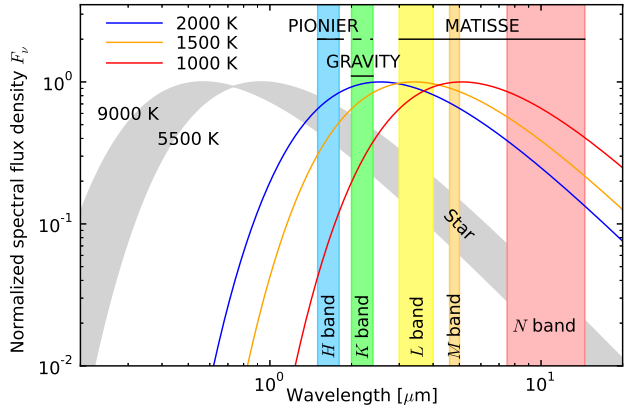}
\caption{Wavelength coverage of the second-generation VLTI instruments (and PIONIER) compared to the
wavelength range in which blackbody dust emission from hot and warm exozodiacal dust peaks. For PIONIER, the
dotted range indicates the K band which is no longer available and the J band which may be reached with a potential
instrument upgrade. \citep{2015Msngr.159...24E}.}
\label{fig:newinstru}
\end{figure}

Future instruments will be designed to answer the most important questions about the origin of these exozodis and 
what they tell us about inner regions of planetary systems. Second generation VLTI instruments such as GRAVITY, MATISSE (together with the current PIONIER) will lead to multiwavelength measurements of the SED over a large range of dust temperatures (see Fig.~\ref{fig:newinstru}). 
Having a more complete SED may reveal a connection between the hot and warm dust. MATISSE will enable us to explore a new range of disc temperatures (from 300 to 1000K) but also to explore dust properties through the potential detection of spectral features (e.g. 3 and 10$\mu$m
silicate features).

Further characterisation, and first directly resolved images of bright exozodis might be soon enabled by second-generation high-contrast imagers working in polarimetric mode in the visible range, such as SPHERE/ZIMPOL or SCExAO/VAMPIRES. 
By combining extreme adaptive optics and polarimetric differential imaging, these instruments can in principle reach very high contrasts at angular separations as small as 20mas from the star, which might be enough to resolve the brightest exozodis, 
or even the sublimation radius in the nearest/widest discs. Such an image would bring invaluable information on the exact location of the dust populations and on their grain properties.

A new more powerful beam combiner called JouFLU has now replaced FLUOR on the CHARA array \citep{2013JAI.....240005S,2017PASP..129b4002N}. Also, new instruments may be built that would enable us to go to the next level. The VLTI currently achieves a dynamic range of a few 10$^{-3} $ in the near-infrared and second-generation instruments are not designed 
to do better. Based on the experience gained with PIONIER, as well as with ground-based mid-infrared nulling instruments (KIN, LBTI), it would be possible to unlock the next level of high-dynamic range observations of the VLTI with a nulling interferometric instrument operating in the thermal 
infrared, a sweet spot to image and characterise young extra-solar planetary systems. With an anticipated dynamic range of 10$^{-4}$, a high contrast thermal IR instrument at the VLTI would be sensitive to faint exozodi emissions around nearby main-sequence stars ($\sim$50 
times the density of the zodiacal cloud) and would inform us on the faint-end of the exozodi luminosity function (complementarity with the LBTI in the Northern hemisphere).

Also, astrometric missions could be used to observe exozodis. It was shown recently that small dust clouds whose IR-excesses could not be detected by current instruments can mimic an Earth-like astrometric signal,
which may affect future astrometric missions looking for exo-Earths \citep{2016A&A...592A..39K}. However, with further observations at different wavelengths,
these dust clouds could be distinguished from a planet. If such dust clouds are detected with future astrometric missions, this would lead to a new complementary way to characterise this dust and increase significantly the sample of known exozodis. 

Many mysteries remain regarding the exact nature and origin of exozodiacal dust. However, the potential for significant progress in the coming decades is good \citep{2017arXiv170308560K}. A development in our understanding of the physics of tiny submicron grains is key. 
A continuation of the current efforts, both theoretical and observational will eventually enable to unravel their origin.

\section{Conclusion}

The aim of this review is to assess our current understanding of exozodis, that is warm or hot dust in the inner regions of planetary systems. Observationally, there is a large sample of exozodis detected, with a dozen systems characterised in detail. {\em Spitzer} and {\em WISE}
have provided tens of mid-IR excesses using spectro-photometry. The real break-through in the last decade, however, has been the interferometric detection and characterisation of {\bf these exozodis (see Table~\ref{tab1})}. The warm dust that probes 
dust located at a few au (Earth's zodi) is detected around $\sim$ 20\% of stars owing to the use of nullers such as the KIN or the LBTI. The hot dust, at a similar location to the Solar System's F-corona, is detected around $>10$\% of stars using visibilities observed
with interferometers such as the CHARA array or the VLTI. Our 
understanding of these observations is now suffciently advanced that we can be confident in the detections. What is known for sure is that between 10 and 30\% of the main sequence stars show a photospheric excess of about 1\% in the near-infrared that is produced within the field-of-view 
of the interferometers (i.e. a maximum of a few au). We are also confident that this dust is dominated by thermal emission and should be hot enough to emit in the H and K bands (i.e. $>$1000K). We also know that the excesses can be variable, which suggests episodic production/replenishment
of dust.
Radiative transfer modelling of these hot exozodis is less direct but lead to 
a few interesting constraints on the dust properties: 1) The dust is located within an au, close to the sublimation rim. 
2) The dust mass of these belts is $\sim$ $10^{-8}-10^{-9}$ M$_\oplus$. 3) Dust grains composing these hot exozodis are of submicron size. 4) The grains seem to be composed of carbonaceous material (rather than pure silicate). Finally, on the theory-side, 
all the new models proposed to create these exozodis are more speculatives and none can fully explain all the observations at once.

However, there are still many unknown concerning these hot systems. Firstly, is it really dust that is observed or some yet unknown phenomenon? This has not yet been proven for sure even though dust is the most likely possibility. 
Also, we do not really know how hot the dust is in these systems and thus how close to the host star it is located and how compact it is. The main unknown is where the dust comes from and what keeps it in place. We know that it cannot be produced locally but must come from the outside. We then either need a
very high replenishment rate or a lower one along with some efficient trapping mechanisms. Finding models that produce both the high replenishment rates needed and the high occurence rate of such systems observed is not easy. Moving the dust all the way from an exo-Kuiper belt to the inner regions is not very efficient \citep[e.g.][]{2012A&A...548A.104B}. 
Exo-asteroid belts may be a more favourable source along with some efficient trapping mechanisms \citep[e.g.][]{2013A&A...555A.146L,Marion}.

Another unknown is the impact of exozodiacal dust on exo-Earth imaging. While several works have studied the impact of certain levels of habitable zone dust on future space missions aiming at imaging and characterising habitable planets, 
the amount of dust present in the habitable zones around nearby stars is still unknown. The first results from the HOSTS survey (Ertel et al., in prep.) are nervously awaited. However, even when the mid-infrared levels of dust are better constrained, the dust properties 
are still unknown. Thus, extrapolating from the thermal dust emission in the mid-infrared to its scattered light brightness in the visible and near-infrared is very uncertain. A better characterisation of the exozodiacal dust through multi-wavelength observations is 
critical to addressing this problem. In addition, the impact of hot exozodiacal dust on exo-Earth imaging is still unclear. While this emission stems from closer to the star and thus does not directly produce noise in the imaging of habitable planets, it 
produces extended emission around the star that will not be blocked by a coronagraph as well as the compact, virtually point-like star. \citet{Kirch17} found that a few percent of the detected near-infrared emission can come from scattered light. The near-infrared disc-to-star 
flux ratio is typically $\sim$ 1\% around $\sim$ 20\% of the nearby stars and the size of this emission is at least a few stellar radii. Assuming gray scattering, this means that 20\% of the nearby stars have extended emission at a contrast of $\sim 10^{-4}$ in the visible. At an intended contrast of 
$10^{-10}$ further out in the habitable zone to be able to detect exo-Earths, this extra emission might pose a serious problem for new missions such as HABEX or LUVOIR.

On an observational level, LBTI, MATISSE will revolutionise the field over the next decade. If the exozodi emission is sufficiently polarised, polarimetric observations using SPHERE/ZIMPOL or SCExAO/VAMPIRES will also lead to further discoveries. 
On a theoretical level our understanding of how the observed levels of hot dust can be sustained has progressed
 from the identification of a problem over a decade ago, to a wealth of theories that aim to explain the high levels of dust, including a link with outer belts, trapping due to magnetic fields or gas, however, none provide the full solution. An improved understanding of the connection between 
hot, warm and cold dust is critical, which will be provided by future multi-wavelength studies. This back and forth between theory and observation will be crucial in the next decade to finally explain the ubiquitous presence of exozodis.

\section*{Acknowledgments}
We thank the referee for an helpful review that improved the quality of the paper. QK and RvL acknowledge support from the European Union through ERC grant number 279973. QK acknowledges funding from STFC via the Institute of Astronomy, Cambridge Consolidated Grant. 
AVK acknowledges support by Deutsche Forschungsgemeinschaft (DFG), grant Kr 2164/15-1. AB acknowledges the support of the Royal Society in terms of a Dorothy Hodgkin Fellowship. 
A large part of the work reviewed in this paper was realised and supported by the French National Research Agency (ANR) through Contract ANR-2010 BLAN-0505-01 (EXOZODI).

\bibliographystyle{plainnat}
\bibliography{references}


\end{document}